\documentclass[twocolumn,amsmath,amssymb, pra]{revtex4}

\usepackage{graphicx}
\usepackage{bm}

\begin{document}

\title{Quantum secret sharing with continuous-variable cluster states}

\author{Hoi-Kwan Lau\footnote{kero.lau@mail.utoronto.ca} and Christian Weedbrook}
\affiliation{Center for Quantum Information and Quantum Control (CQIQC),
 Department of Physics, University of Toronto, 60 Saint George Street, Toronto, M5S 1A7 Ontario, Canada}

\date{\today}

\begin{abstract}
We extend the formalism of cluster state quantum secret sharing, as presented in Markham and Sanders [Phys. Rev. A \textbf{78}, 042309 (2008)] and Keet \textit{et al.} [Phy. Rev. A \textbf{82}, 062315 (2010)], to the continuous-variable regime.  We show that both classical and quantum information can be shared by distributing continuous-variable cluster states through either public or private channels.  We find that the adversary structure is completely denied from the secret if the cluster state is infinitely squeezed, but some secret information would be leaked if a realistic finitely squeezed state is employed.  We suggest benchmarks to evaluate the security in the finitely squeezed cases.  For the sharing of classical secrets, we borrow techniques from the continuous-variable quantum key distribution to compute the secret sharing rate.  For the sharing of quantum states, we estimate the amount of entanglement distilled for teleportation from each cluster state.
\end{abstract}

\pacs{03.67.Dd, 03.67.Ac, 03.67.Hk}

\maketitle

\section{Introduction}

Secret sharing is a cryptographic task aiming to distribute a secret amongst a group of parties.  A good secret sharing protocols should allow authorised subsets of parties, known as the \textit{access structure}, to faithfully reconstruct the secret, while other unauthorized parties, known as the \textit{adversary structure}, are denied any information about the secret.  Classical secret sharing protocols have been proposed \cite{Blakley:1979p14072, Shamir:1979p14071} where classical information is encoded by a mathematical transformation.  The protocols can be proven to be information-theoretically secure, i.e., no information about the secret can be obtained by the adversary structure even when they have unlimited computational power, if the communication channels between the dealer and the parties are secure.

Following the rapid development of quantum information, the extension of secret sharing to the quantum regime has received much theoretical attention \cite{Hillery:1999p5705, Cleve:1999p5862, Gottesman:2000p13957, Tyc:2002p13582}.  The objective of quantum secret sharing (QSS) is to use the quantum correlations in well-constructed entangled states to securely transmit a set of classical or quantum information to only the access structures.  As the involving parties are supposed to be spatially well separated, an optical system is the most suitable implementation for QSS due to its excellent mobility.  Several proof-of-principle experiments have already been demonstrated \cite{Tittel:2001p14241, Chen:2005p14244, Lance:2004p13317}.  However, constructing a large-scale optical QSS state is technically challenging, because the nonlinear interaction between photons is weak, and some QSS protocols require more than two quantum levels, where the commonly employed polarisation encoding is not applicable.  Recently, Markham and Sanders \cite{Markham:2008p13316} proposed a unified QSS approach based on qubit cluster states \cite{Raussendorf:2001p3202, Raussendorf:2003p2988}, which could be constructed efficiently using only linear optics and post-selection \cite{Nielsen:2004p8340, Browne:2005p10573, Duan:2005p4737}.

Cluster states have another advantage in that an $N$-mode cluster is well characterised by $N$ stabilisers or an $N$-vertex connected graph, in contrast to a general quantum state that have to be expressed in an exponential number of superpositions.  Therefore, the theoretical construction and the security analysis of a cluster state QSS scheme could be simplified.  The idea of cluster state QSS has also been extended to odd-dimensional states (qudits) in \cite{Keet:2010p13444}.  

In this paper, we further extend the idea of cluster state QSS into the continuous-variable (CV) regime.  While many quantum information protocols can be optically implemented by using discrete- or continuous-variable formalism, CV systems have the advantages that multi-partite entangled states can be produced deterministically, and the measurement is high in fidelity using present technology.  In particular, CV cluster states are proposed to be useful resources to conduct measurement-based universal quantum computation \cite{Menicucci:2006p6585, Weedbrook:2012p13102}.  A CV cluster state can be efficiently implemented in an optical system by various approaches including the conventional method of controlled-phase (\small CPHASE\normalsize) operation \cite{Menicucci:2006p6585,Zhang:2006p14445}, linear optics with offline squeezing \cite{VanLoock:2007p13575}, optical parametric oscillator \cite{Menicucci:2007p14440, Menicucci:2008p14441, Flammia:2009p14442}, and quantum nondemolition gate \cite{Menicucci:2010p14443}.  Recently, CV cluster states involving four optical modes have been demonstrated experimentally \cite{Su:2007p14284, Yukawa:2008p14285}.  For simplicity, we consider the CV cluster states are prepared by the conventional method of \small CPHASE \normalsize operation, though the states can be equivalently prepared by other approaches, and our result is independent to the method of state preparation.

The main objective of this work is to investigate how CV cluster states can be used to securely share quantum and classical secrets.  
Instead of directly extending the qudit approach to the $d \rightarrow \infty$ limit, a CV cluster state is critically different from its discrete-variable counterparts in that a perfect (infinitely squeezed) CV cluster state is physically and hence, practically, impossible.  
We find that when realistic finitely squeezed cluster states are instead utilised, QSS is still possible but the security is inevitably reduced, i.e., the secret is not precisely recovered by the access structure while partial information is leaked to unauthorised parties.  We suggest benchmarks to evaluate the performance of each of the QSS tasks.  For the sharing of classical information, we calculate the amount of secure key that can be distilled from each cluster state for encoding the secret.  A procedure is provided to transform the distilled state to the standard form that can be analysed by the techniques in CV quantum key distribution (QKD).  For the sharing of quantum information, we estimate the number of cluster states required to establish a high fidelity teleportation channel to transmit the secret state.  The amount of entanglement is quantified by the logarithmic negativity.  In both tasks, we give two examples to demonstrate the decoding and security analysis procedures.

As we want to focus our discussion on the application of the quantum correlations of CV cluster states, the states received by the parties are assumed to be the same as when prepared by the dealer, i.e., all quantum channels are ideal (noiseless and lossless).  Detections are also assumed to be perfect in fidelity.  

Our paper is outlined as follows.  In Sec. \ref{sec:background}, we introduce QSS and classify it into three tasks.  The physical and mathematical background of CV cluster states are also reviewed.  In Sec. \ref{sec:CC QSS} and \ref{sec:CQ QSS}, we analyse the security of classical information sharing when the cluster state is delivered through secure and insecure channels respectively.  In Sec. \ref{sec:QQ QSS}, we discuss the performance of quantum state sharing.  We conclude in Sec \ref{sec:conclusion}. with a short discussion.  

We denote the quantities of the access structure by the subscript $A$, that of the adversary structure by $E$, and that of the dealer by $D$.  We pick $\hbar=1$ in the following calculations, and all logarithms are to base 2.

\section{Background \label{sec:background}}

\subsection{Quantum secret sharing \label{sec:QSS}}

In literatures, the idea of QSS is developed to serve one of the following three tasks \cite{Markham:2008p13316}:

CC:  Classical information is shared among parties by distributing QSS states through private (secure) channels, which are invulnerable to eavesdropping.  The role of quantum resources is to substitute the classical secret sharing encoding by the quantum correlations in a QSS state.

CQ: Classical information is shared among parties by distributing QSS states through public (insecure) channels, which are open for eavesdropping.  The quantum correlations in the QSS states can be used both to detect the disturbance of eavesdropping and to encode secret sharing.  When comparing with the hybrid approach that unifies classical secret sharing and QKD, the CQ scheme can reduce the cost of communication \cite{Hillery:1999p5705}.

QQ: Also known as quantum state sharing, a secret quantum state is shared among parties by distributing QSS states through public channels.  The QQ scheme can be implemented by either encoding the quantum secret into a QSS state, or using a QSS state to distribute entanglement between the dealer and the access structure for teleporting the secret state.  We consider the later approach in the current paper.

The three tasks form a hierarchy of the required resources, i.e., a QQ quantum state can perform all the three tasks, and a CQ state can be used for CC, while the reverse is not always true.  In principle, constructing a QQ state is versatile, but the amount of resources and the required infrastructure can be optimised according to the properties of the shared information and the channels.

For CC and CQ, we consider the cluster states are measured by the dealer and the access structure.  Because of the entanglement, random but strongly correlated measurement outcomes will be obtained, from which the dealer and the access structure can distill secure keys.  Therefore the secret sharing rate, i.e., the amount of classical information securely shared in each round of QSS, is determined by the net amount of secure key distilled from each cluster state.

For QQ, we consider the dealer and the access structure extract entanglement from the cluster states.  After accumulating enough extracted states, entanglement distillation can be conducted to distill a more entangled state, through which the secret state can be teleported from the dealer to the access structure with higher fidelity.

We note that in all the QSS tasks, the objective of the dealer is to securely transmit the secret to the access structure, although the identities of the access structure are not revealed until all QSS states have been received.  Because, in a secure protocol, the mutual information between the dealer and the access structure is larger than the information obtained by the adversary structure, the access structure's identities can be authenticated using parts of the shared information.  The dealer should then trust the access structure and co-operate in subsequent post-processing of the shared QSS states.

We also note that in the limit of infinite squeezing, our cluster state scheme is not as general as the QQ scheme proposed in Ref. \cite{Tyc:2002p13582}.  However our scheme is interesting because all three kinds of QSS are considered in a unified approach, and the resource state is a cluster state that can be efficiently constructed and can be easily analysed.

\subsection{Continuous-variable cluster states \label{sec:cluster}}

As an analog to the discrete-variable cluster state, which is formed by preparing all qudits in an eigenstate of the generalised Pauli $X$ operator and then applying \small CPHASE\normalsize gate, a CV cluster state is formed by first preparing all quantum modes as squeezed vacuum states and applying CV \small CPHASE \normalsize gates, given by $\hat{C} = \exp\{i \mathcal{A}_{ij}\hat{q}_i \hat{q}_j \} $.  An $n$-mode CV cluster state can be characterised by an $n$-vertices graph, where the quantum modes act as the vertices $\mathcal{V}=\{v_i \}$, and a \small CPHASE \normalsize operation is applied across edges $\mathcal{E}=\{e_{ij} = \{v_i, v_j \} \} $ with weight $\mathcal{A}_{ij}$ \cite{Keet:2010p13444}.  The CV cluster state $|\Psi\rangle$ is defined as
\begin{equation}
|\Psi\rangle := \prod_{e_{ij}\in \mathcal{E}} \exp\{i \mathcal{A}_{ij}\hat{q}_i \hat{q}_j \} |\psi_0\rangle^{\otimes n} ~,
\end{equation}
where in the infinitely squeezing case
\begin{equation}
|\psi_0\rangle_{\textrm{infinite}} = |0\rangle_p~,~~\textrm{where}~~\hat{p} |0\rangle_p=0~,
\end{equation}
and in the finitely squeezing case 
\begin{equation}
|\psi_0\rangle_{\textrm{finite}}  = \frac{\sqrt{\sigma}}{\pi^{1/4}} \int e^{-\sigma^2 q^2/2 } |q\rangle_q dq~,
\end{equation}
where $|q\rangle_q$ is the eigenstate of $\hat{q}$ with eigenvalue $q$; $\sigma$ is a parameter characterising the degree of squeezing.

\subsubsection{Nullifier representation}

Apart from the ket vector representation, an infinitely squeezed cluster state can be characterised by its stabilisers \cite{GottesmanThesis, Weedbrook:2012p13102}.  A stabiliser $\hat{S}$ of a state $|\psi\rangle$ is defined as the operator of which $|\psi\rangle$ is an eigenstate with +1 eigenvalue, i.e., $\hat{S}|\psi\rangle = |\psi\rangle$.  Analogous to the discrete-variable cluster state, an $n$-mode infinitely squeezed CV cluster state has at least $n$ independent stabilisers.  Although any sum and product of the stabilisers is a new stabiliser, the whole set of stabilisers uniquely specifies the cluster state \cite{Gu:2009p12232}.  

In a CV system, considering the nullifiers of a cluster state is sometimes more convenient than the stabilisers.  A nullifier $\hat{N}$ is defined as an operator of which $|\psi\rangle$ is an eigenstate with eigenvalue 0, i.e., $\hat{N}|\psi\rangle=0$.  There can be infinitely many choice of nullifiers as any sum and product of nullifiers is another nullifier.  For an infinitely squeezed CV cluster state, we choose a set of nullifiers, which we call the standard set, i.e., \cite{Gu:2009p12232}
\begin{equation}\label{eq:nullifier}
\hat{N}_i = \hat{p}_i - \sum_{j \in \mathcal{N}} \mathcal{A}_{ij} \hat{q}_j~,
\end{equation}
where the position operators are summed over indices of the neighbours of the vertex $i$ in the graph, i.e., ${j|(i,j)\in \mathcal{E}}$.  The standard nullifiers can be constructed by considering that before the \small CPHASE \normalsize operations, the squeezed vacuum modes are nullified by $\hat{p}_i$'s.  The \small CPHASE \normalsize operation between the mode $i$ and $j$ transforms the nullifiers as $\hat{p}_i \rightarrow e^{i \mathcal{A}_{ij}\hat{q}_i \hat{q}_j }\hat{p}_i e^{-i \mathcal{A}_{ij}\hat{q}_i \hat{q}_j }=\hat{p}_i - \mathcal{A}_{ij}\hat{q}_j$.  From the construction procedure, it can be easily shown that all standard nullifiers commute and are linearly independent.



\subsubsection{Wigner function representation}

As an extension to the nullifier representation, the Wigner function is a good description of the quantum correlation of finitely squeezed CV cluster states.  
The Wigner function of a single mode CV state $\hat{\rho}$ is defined as \cite{Gu:2009p12232}
\begin{equation}
W(q,p) := \frac{1}{2 \pi} \int_{-\infty}^{\infty} \exp(ipx) \left\langle q-\frac{x}{2}\right|_q\hat{\rho}\left| q+\frac{x}{2} \right\rangle_q dx~,
\end{equation}
where the definition can be trivially generalised to the multi-mode state.  The Wigner function of $n$ finitely squeezed vacuum states is given by
\begin{equation}
W_0(\bm{q},\bm{p}) = \frac{1}{\pi^n}\prod_i^n \exp(-\sigma_i^2 q_i^2) \exp\left(-\frac{p_i^2}{\sigma_i^2}\right),
\end{equation}
and that of a finitely squeezed CV cluster state is
\begin{equation}\label{eq:Wigner}
W_{c}(\bm{q},\bm{p}) \equiv W_0(\bm{q},\bm{N}) = \frac{1}{\pi^n}\prod_i^n \exp(-\sigma_i^2 q_i^2) \exp\left(-\frac{N_i^2}{\sigma_i^2}\right)~,
\end{equation}
where  $\bm{q}=(q_1,\ldots q_n)^T$, $\bm{p}=(p_1,\ldots p_n)^T$, and $\bm{N}=(N_1,\ldots N_n)^T$; $N_i$ is the standard nullifier in Eq.~(\ref{eq:nullifier}) with the operators replaced by the respective scalar variables;  the initial degree of squeezing of each mode $i$ is $\sigma_i$. In the infinitely squeezing limit, i.e., $\sigma_i\rightarrow 0 ~ \forall i$, the $\exp(-\sigma_i^2 q_i^2)$ would converge to a constant function while the $\exp(-N_i /\sigma_i^2)$ term becomes a delta function, i.e.,
\begin{equation}\label{eq:WignerDelta}
W_{\textrm{infinite}}(\bm{q},\bm{p}) \propto \prod_i^n\delta(N_i) ~.
\end{equation}

\subsubsection{Correlations of measurement \label{subsec:correlation}}

Consider an infinitely squeezed CV cluster state is locally measured by the operators $\{\hat{M}_i\}$, where $\hat{M}_i$ is a linear combination of $\hat{q}_i$ and $\hat{p}_i$, i.e., homodyne detection in a rotated basis.  If the measurements are compatible to a nullifier, there exists a linear combination of $\hat{M}_i$'s that equals to a linear combination of standard nullifiers, i.e., $\sum_{i=1,n}k_i \hat{M}_i = \sum_{i=1,n}l_i \hat{N}_i$ for some real $k_i$'s and real $l_i$'s, 
then the measurement outcomes would be correlated due to the entanglement as $\sum_{i=1,n} k_i M_i =0$.

Similar quantum correlations of measurements prevail in finitely squeezed CV cluster states, but the accuracy depends on the degree of squeezing.  Consider a finitely squeezed CV cluster state is measured by the same set of measurement operators $\{\hat{M}_i\}$, the expectation value of the measurement outcomes are statistically correlated as in the infinitely squeezed case, i.e.,
\begin{equation}
\Big\langle \sum_{i=1,n}k_i \hat{M}_i \Big\rangle =  \Big\langle \sum_{i=1,n}l_i \hat{N}_i \Big\rangle =0 ~.
\end{equation}
However, the variance is finite, i.e.,
\begin{eqnarray}
&&\Big\langle \Delta\Big(\sum_{i=1,n}k_i \hat{M}_i \Big)^2 \Big\rangle = \Big\langle \Delta\Big(\sum_{i=1,n}l_i \hat{N}_i \Big)^2 \Big\rangle \nonumber \\
&=&\int \Big(\sum_{i=1,n} l_i N_i\Big)^2 W_c(\bm{q},\bm{p}) d^n\bm{q} d^n\bm{p} = 
 \sum_{i=1,n} \frac{l_i^2 \sigma_i^2}{2}~,\label{eq:strong}
\end{eqnarray}
but scales as $\sigma_i^2$ that is small.  The correlation comes from the $\exp(-N_i^2/\sigma_i^2)$ terms in Eq.~(\ref{eq:Wigner}), which are narrow width Gaussian functions.  

In subsequent discussions, we regard a quantum correlation is ``strong" if the collective variance of the local measurement outcomes is small, and so the modes are strongly correlated if their local operators produce strong correlations.  Our QSS scheme is secure if the access structure is stronger correlated to the secret than the unauthorised parties.  According to Eq.~(\ref{eq:strong}), the local measurement operators are strongly correlated if they linearly combine as a nullifier and $\sigma_i$'s are small, so we usually encode the secret in nullifiers.

\subsubsection{Cluster-class state}

A class of states that shares similar properties as the CV cluster state can be constructed by applying local Gaussian operators onto $|\Psi\rangle$.  The operations linearly transform the quadrature operators in nullifiers, as well as the quadrature parameters in the Wigner function, as $\hat{q}\rightarrow a_q \hat{q} + b_q \hat{p} +c_q$ and $\hat{p}\rightarrow a_p \hat{q} + b_p \hat{p} +c_p$ for some real constants $a,b,c$ that obey the uncertainty principle.  General linear transformations can be implemented by only three kinds of basic operators \cite{Lloyd:1999p11945}: displacement, squeezing, and Fourier operator.  

A displacement operator $\hat{D}(\alpha)$ shifts a nullifier by a constant factor, i.e., the components in nullifiers are transformed as $\hat{q} \rightarrow \hat{q} + \sqrt{2} \textrm{Re}(\alpha)$ and $\hat{p} \rightarrow \hat{p} + \sqrt{2}\textrm{Im}(\alpha)$.  
All the displacements do not affect the measurement basis nor the variance of the quantum correlations, but only the expectation values of the measurement results are changed. A squeezing operator $\hat{S}(\gamma)=\exp(-ir(\hat{q}\hat{p}+\hat{p}\hat{q})/2)$ scales the quadrature operators as $\hat{S}^\dag \hat{q} \hat{S} \rightarrow \gamma \hat{q}$ and $\hat{S}^\dag \hat{p} \hat{S} \rightarrow \hat{p}/\gamma$, where $\gamma=e^r$.  
Linear coefficients of $\hat{x}_i$ and $\hat{p}_i$ in the nullifiers will be altered but the measurement basis remains the same. A Fourier operator $\hat{F}(\theta)=\exp(-i\theta(\hat{q}^2+\hat{p}^2)/2)$ transforms the quadrature operators as $\hat{F}^\dag \hat{q} \hat{F} = \cos \theta \hat{q} + \sin \theta \hat{p}$ and $\hat{F}^\dag \hat{p} \hat{F} = -\sin \theta \hat{q} + \cos \theta \hat{p}$.  
The Fourier operator changes the local measurement bases for the quantum correlation.


\section{CC Quantum secret sharing \label{sec:CC QSS}}

In the CC setting of QSS, the dealer is connected to the $n$ parties through secure quantum channels.  
A classical secret value $s$ is encoded by displacing certain modes $i$ of the cluster state by some function $f_i(s)$.
The value of $f_i(s)$, the strength of the \small CPHASE \normalsize $\mathcal{A}_{ij}$, and the neighbours of the cluster $\mathcal{N}$ are designed for specific access and adversary structure.
A CV cluster state can be used for CC QSS if for each access structure, there is a nullifier containing $s$ and the local quadrature operators of only that access structure, i.e., there exists real numbers $l_i$ such that
\begin{equation}
\sum_i^N l_i \hat{N}_i = \sum_{j \in A} k_j \hat{M}_j + g(s)~,
\end{equation}
where $k_j$ are real numbers; $\hat{M}_j$ are linear combinations of the local quadrature operators of an access structure party; $g(s)$ is a nontrivial function of $s$.  On the other hand, every adversary structure cannot construct a nullifier containing $s$ and only their local operators.


In the case of infinitely squeezing, the access structure can obtain $g(s)$, and thus $s$, by locally measuring their modes according to $\hat{M}_j$. The scheme is secure if the reduced Wigner function of the adversary structure is independent of $s$.  

In the case of finitely squeezing, the access structure also measures according to $\hat{M}_j$.  Their results are merely strongly correlated to $s$, while some information about the secret is leaked to the adversary structure due to the weak correlation.
The security of the QSS scheme can be analysed by comparing the amount of information obtained by the access structure and the adversary structure.

The information obtained by the access structure is quantified by the mutual information, $I(D:A)$, between the dealer and the access structure \cite{ANielsen:2000p5658}.  Let the dealer chooses a secret value $s$ following a probability $\mathcal{P}_D(s)$.  The access structure would not obtain exactly the same value due to the finite squeezing.  The probability of obtaining a result $s'$ follows a probability distribution $\mathcal{P}_{A|D}(s,s')$, while the total probability, $\mathcal{P}_A(s')$, of the access structure's result is given by
\begin{equation}
\mathcal{P}_A(s')=\int \mathcal{P}_D(s) \mathcal{P}_{A|D}(s,s')ds~.
\end{equation}
The mutual information $I(D:A)$ is defined as \cite{ANielsen:2000p5658}
\begin{equation}\label{eq:mutualcc}
I(D:A) = H(A)-H(A|D)~,
\end{equation}
where $H(A)$ is the entropy of the access structure's result, which is defined as
\begin{equation}
H(A) = -\int \mathcal{P}_A(s') \log \mathcal{P}_A(s') ds'~,
\end{equation}
and $H(A|D)$ is the entropy of access structure conditioned on knowing $s$, which is defined as \cite{ANielsen:2000p5658}
\begin{equation}
H(A|D) = -\int \mathcal{P}_D(s) \mathcal{P}_{A|D}(s,s') \log \mathcal{P}_{A|D}(s,s')dsds'~.
\end{equation}

On the other hand, the adversary structure can unite their modes through ideal quantum channels, and can conduct any operation allowed by physics.  The amount of information leaked to the adversary structure, $I(D:E)$, is capped by the Holevo bound $\chi$ \cite{ANielsen:2000p5658,Holevo:1973p14339}, i.e.,
\begin{equation}
I(D:E) \leq \chi = S(\hat{\rho}_E)-\int \mathcal{P}_D(s) S(\hat{\rho}_{E|D}(s)) ds~,
\end{equation}
where $S(\hat{\rho})$ is the von Neumann entropy; $\hat{\rho}_{E|D}(s)$ is the state obtained by the adversary structure if $s$ is prepared by the dealer;  $\hat{\rho}_E$ is the average state obtained by the adversary structure, viz.
\begin{equation}
\hat{\rho}_E = \int \mathcal{P}_D(s) \hat{\rho}_{E|D}(s) ds~.
\end{equation}

As CV cluster states are Gaussian, the reduced state of the adversary structure is also Gaussian.  The von Neumann entropy of Gaussian states can be calculated by using their covariance matrix $\bm{V}$, which is defined as $V_{ij} := \langle\{\Delta x_i, \Delta x_j\}\rangle/2$ \cite{Weedbrook:2012p13102}.   If the Wigner function of an $r$-mode Gaussian state is known, $\bm{V}$ can be obtained through the relation \cite{Weedbrook:2012p13102}
\begin{equation}
W(\hat{x}) = \frac{\exp(-1/2 (\bm{x}-\bar{\bm{x}})^T \bm{V}^{-1} (\bm{x}-\bar{\bm{x}}))}{(2\pi)^r \sqrt{\det\bm{V}}}~,
\end{equation}
where $\bm{x}=(q_1,p_1,\ldots,q_r,p_r)^T$.  Covariance matrices can be characterised by their symplectic spectrum $\{\nu_k\}$, which is equal to the eigenspectrum of the matrix $|i\bm{\Omega}\bm{V}|$ \cite{Weedbrook:2012p13102}, where
\begin{eqnarray}
\Omega_{i,j}= \Big \{ \begin{array}{lll}
1 & \textrm{ if }~i=2k-1,&j=2k~,\\
-1 & \textrm{ if }~i=2k,&j=2k-1~,\\
0 & \textrm{else,}
\end{array}
\end{eqnarray}
for $k=1,\ldots, r$.  The von Neumann entropy is calculated using
\begin{equation}\label{eq:Neumann}
S(\hat{\rho})=\sum_i^r g(\nu_k)~,
\end{equation}
where
\begin{equation}
g(\nu):=\left(\nu+\frac{1}{2}\right)\log\left(\nu+\frac{1}{2}\right)-\left(\nu-\frac{1}{2}\right)\log\left(\nu-\frac{1}{2}\right)~.
\end{equation}

In the case that the covariance matrices of $\hat{\rho}_E$ and $\hat{\rho}_{E|D}$ are independent of $s$, their respective von Neumann entropies are also so.  Then the Holevo bound can be simplified as,
\begin{equation}\label{eq:simHolevo}
I(D:E)\leq S(\hat{\rho}_E)-S(\hat{\rho}_{E|D})~.
\end{equation}
The minimum secret sharing rate in each round of the protocol is thus 
\begin{equation}\label{eq:Kcc}
K_{cc}=I(D:A)-I(D:E)~.  
\end{equation}
With the strongly correlated random numbers $s$ and $s'$, and the expected amount of secret information, $K_{cc}$, a secure key can be distilled to encode and share the classical secret \cite{VanAssche:2004p13958}.

As examples, $K_{cc}$ for the (2,3)- and the (3,5)-protocol are calculated.  Readers who are mainly interested in the general formalism of QSS can skip the examples.


\subsection{Example 1 of CC Quantum secret sharing: (2,3)-protocol}

\begin{figure}
\centering
\includegraphics{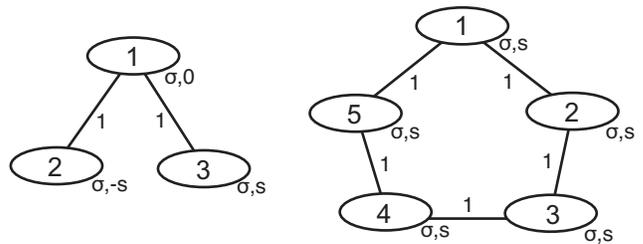}\caption{\label{fig:CClayout} Schematic representation of the cluster states for the (2,3)-protocol (left), and the (3,5)-protocol (right).  Each oval represents a squeezed mode, which will be distributed to the party, denoted by the label inside the ovals.  The subscript of each mode denotes the squeezing parameter and the displacement before the \small CPHASE \normalsize operation.  The edges joining modes represent \small CPHASE \normalsize operations with the strength denoted by the edges' label.}
\end{figure}

In a (2,3)-CC protocol, the access structure is any two of the three parties collaborating, while the adversary structure is any collaboration with only one party.  The (2,3)-CC protocol can be implemented by a linear three-mode cluster, as shown in Fig.~\ref{fig:CClayout}.  We assume the dealer picks the secret classical value $s$ following a Gaussian probability distribution with a width $\Sigma$, i.e.,
\begin{equation}\label{eq:PD}
\mathcal{P}_D(s) = \frac{1}{\sqrt{\pi}\Sigma} e^{-s^2/\Sigma^2}~.
\end{equation}
The state is encoded by displacing mode $2$ by $is/\sqrt{2}$ and mode $3$ by $-is/\sqrt{2}$.  In the infinitely squeezing case, the nullifiers of the CV cluster state are
\begin{equation}\label{eq:cc23}
\hat{N}_1 = \hat{p}_1-\hat{q}_2-\hat{q}_3~;~\hat{N}_2 =\hat{p}_2-\hat{q}_1+s~;~\hat{N}_3 =\hat{p}_3-\hat{q}_1-s~.
\end{equation}

\subsubsection{Parties \{1, 2\} collaborating}

Consider parties $1$ and $2$ are the access structure.  If the cluster state is infinitely squeezed, the shared secret is the difference between $\hat{p}$ measurement outcome of party $1$ and $\hat{q}$ measurement outcome of party $2$, i.e., $s=q_1-p_2$ according to $\hat{N}_2$.  Deriving from Eq.~(\ref{eq:WignerDelta}), the reduced Wigner function of party $3$ is a constant function independent of $s$, thus the protocol is secure. In the finitely squeezing case, the Wigner function is given by Eq.~(\ref{eq:Wigner}) with the nullifiers in Eq.~(\ref{eq:cc23}).  For simplicity, we assume the modes are equally squeezed, i.e., $\sigma_i = \sigma$ for all $i$, but our analysis is applicable for inhomogeneous $\sigma_i$.

The measurement basis of the parties is the same as the infinitely squeezing case.  The classical probability of the measurement results $p_2$ and $q_1$ is
\begin{equation}
\mathcal{P}^{(s)}_{A|D;\{1,2\}}(q_1,p_2) = \frac{1}{\pi} e^{-(p_2 -q_1 - s)^2/\sigma^2} e^{-\sigma^2 q_1^2}~,
\end{equation}
which is obtained by tracing out $q_1$, $p_2$, and the contributions of party $3$ in the Wigner function.
The probability distribution of the difference of the measurements, $s'=q_1-p_2$, can be obtained by tracing out an orthogonal quantity, e.g. $(p_1+q_2)/2$, then we get
\begin{equation}
\mathcal{P}_{A|D; \{1,2\}}(s,s') = \frac{1}{ \sqrt{\pi}\sigma} \exp\Big(-\frac{(s-s')^2}{\sigma^2} \Big)~,
\end{equation}
and thus
\begin{equation}
\mathcal{P}_{A;\{1,2\}}(s') = \frac{1}{\sqrt{\pi}\sqrt{\sigma^2 +\Sigma^2}} e^{-s'^2/(\sigma^2+\Sigma^2)}~.
\end{equation}
The mutual information between the dealer and the access structure can then be calculated using Eq.~(\ref{eq:mutualcc}).

We now consider the adversary structure.  The reduced Wigner function of $\hat{\rho}_E$ and $\hat{\rho}_{E|D}(s)$ are
\begin{eqnarray}
W_{E|D;\{3\}} &=& \frac{\sigma^2}{\pi \sqrt{1+\sigma^4}} e^{-\frac{\sigma^2 ((p_3-s)^2 + q_3^2 (1+\sigma^4))}{1+\sigma^4}}~; \\
W_{E;\{3\}}&=&\int W_{E|D;\{3\}} ds \nonumber \\
&=& \frac{\sigma^2}{\pi \sqrt{1+\sigma^4+\sigma^2 \Sigma^2}} e^{-\sigma^2 \left(q_3^2+\frac{p_3^2}{1+\sigma^4 +\sigma^2 \Sigma^2} \right) }~.
\end{eqnarray}
The covariance matrices of these states are
\begin{equation}
\bm{V}_{E|D;\{3\}}=\left(\begin{array}{cc}\frac{1}{2\sigma^2} & 0 \\0 & \frac{1+\sigma^4}{2\sigma^2}\end{array}\right)~,~\bm{V}_{E;\{3\}}=\left(\begin{array}{cc}\frac{1}{2\sigma^2} & 0 \\0 & \frac{1+\sigma^4+\sigma^2\Sigma^2}{2\sigma^2}\end{array}\right),
\end{equation}
where the symplectic eigenvalues are  $\nu_{E|D;\{3\}}=\sqrt{1+\sigma^4}/2\sigma^2$ and $\nu_{E;\{3\}}=\sqrt{1+\sigma^4+\sigma^2\Sigma^2}/2\sigma^2$, respectively.  The Holevo bound can then be calculated using Eq.~(\ref{eq:Neumann}) and (\ref{eq:simHolevo}), and hence the secret sharing rate can be obtained from Eq.~(\ref{eq:Kcc}).

Because both party $2$ and $3$ hold the end mode of the cluster state, the structure of their local states is the same, i.e., all Wigner functions will be the same as above except replacing the subscript $2$ by $3$ and $s$ by $-s$.  The security for party $\{1,3\}$ collaboration can be analysed by similar procedure as the $\{1,2\}$ collaboration, and the secret sharing in both collaborations will be the same.

\subsubsection{Parties \{2,3\} collaborating}

Consider parties $2$ and $3$ are now the access structure.  In the infinitely squeezed case, because the operator $\hat{N}_2-\hat{N}_3 = \hat{p}_2-\hat{p}_3+2s$ is also a nullifier, the secret $s$ can be obtained if both parties conduct $\hat{p}$ measurement, i.e., $s=(-p_2+p_3)/2$.  The protocol is secure because the reduced Wigner function of party $1$ is a constant independent of $s$.

In the finitely squeezed case, the measurement results of $\hat{p}_1$ and $\hat{p}_2$ follow a probability distribution
\begin{eqnarray}
\mathcal{P}^{(s)}_{A|D;\{2,3\}}(p_1,p_2) &=& \frac{1}{\pi \sqrt{2+\sigma^4}} \exp\Big(-\frac{(p_2-p_3+2s)^2}{\sigma^2(2+\sigma^4)} \Big)\nonumber \\
&& \times e^{- \sigma^2((p_2+s)^2+(p_3-s)^2)/(2+\sigma^4)}~.
\end{eqnarray}
The first exponent accounts for the strong correlations while the last exponent is higher order weak correlations.  The quantity $s'=(-p_2+p_3)/2$ follows the probability
\begin{equation}
\mathcal{P}_{A|D;\{2,3\}}(s,s') = \frac{1}{\sqrt{2\pi}\sigma} \exp\Big(-\frac{2(s -s')^2}{ \sigma^2} \Big)~,
\end{equation}
and thus
\begin{equation}
\mathcal{P}_{A;\{2,3\}}(s')=\sqrt{\frac{2}{\pi (\sigma^2+2 \Sigma^2)}} \exp\Big(-\frac{2 s'^2}{\sigma^2 +2 \Sigma^2} \Big)~.
\end{equation}

For the adversary structure party $1$, the reduced Wigner function of $\hat{\rho}_{E|D;\{1\}}$ is
\begin{equation}\label{eq:WED1}
W_{E|D;\{1\}} = \frac{\sigma^2}{\pi \sqrt{2+\sigma^4}} e^{-\sigma^2\left(q_1^2+\frac{p_1^2}{2+\sigma^4}\right) }~.
\end{equation}
Because Eq.~(\ref{eq:WED1}) is independent of $s$, the Wigner function of $\hat{\rho}_{E;\{1\}}$ and $\hat{\rho}_{E|D;\{1\}}$ would be the same, i.e., $W_{E|D;\{1\}}=W_{E;\{1\}}$.  Therefore the Holevo bound vanishes and party $1$ cannot get any information, and hence the secret sharing rate is simply $I(D:A)$.  

\subsection{Example 2 of CC Quantum secret sharing: (3,5)-protocol}

In a (3,5)-CC protocol, the access structure is any three of the five parties collaborating, while the adversary structure is any collaboration with less than three parties.  The (3,5)-CC protocol can be implemented by a star-shaped five mode cluster, as shown in Fig.~\ref{fig:CClayout}.
All five modes of the cluster state are displaced by $-is/\sqrt{2}$, where the classical secret $s$ is assumed to be chosen according to the probability distribution given in Eq.~(\ref{eq:PD}).  In the infinitely squeezing case, the nullifiers of the CV cluster state are given by
\begin{eqnarray}\label{eq:cc35}
\hat{N}_i = \hat{p}_i-\hat{q}_{i+1}-\hat{q}_{i-1}-s~,
\end{eqnarray}
where $i+1=1$ when $i=5$; $i-1=5$ when $i=1$.  

Ten different combinations of access structure can be formed in this protocol, but they can be categorised into two classes of collaborations, which are three neighbouring parties, and two neighbours with one disjoint party.  Without loss of generality, we consider parties \{1,2,3\} as an example of three neighbours collaboration, and parties \{1,3,4\} for two neighbours collaboration.  The security proof and secret sharing rate of these two cases can be adapted to other collaborations after indices changing.  

\subsubsection{Parties \{1,2,3\} collaborating}

Consider parties 1, 2, and 3 are the access structure.  In the infinitely squeezing case, the secret $s$ can be obtained if both parties 1 and 3 measure $\hat{q}$ and party 2 measures $\hat{p}$, i.e., $s=-q_1+p_2-q_3$ according to $\hat{N}_2$ in Eq.~(\ref{eq:cc35}).  The reduced Wigner function of parties \{4, 5\} collaboration is a constant function after tracing out the contributions of the access structure in Eq.~(\ref{eq:WignerDelta}).

In the finitely squeezing case, the probability measurement by parties \{1,2,3\} is determined by the reduced Wigner function $W_{A|D;\{1,2,3\}}(q_1,p_1,q_2,p_2,q_3,p_3)$ after tracing out the contributions of parties 4 and 5.  As the measurement bases are $\hat{q}_2$, $\hat{p}_1$, and $\hat{q}_3$, the measurement probability is obtained by tracing out the dependence of $p_1$, $q_2$, and $p_3$ from the reduced Wigner function.  
The probability distribution of the received secret, $s'=p_2 - q_1-q_3$, can be obtained by first substituting the set of variables $(q_1,p_2,q_3)$ by another linearly independent set of variables $(q_1, s', q_3 )$, and then tracing out $q_1$ and $q_3$.  The Jacobian matrix of this variable transformation is $1$, so the form of probability distribution remains the same \cite{ProbBook}.

The tracing processes described above involves different physical meanings, but the end result is that all contributions except $s'$ are traced out.  So the probability distribution of $s'$ can be obtained in only two steps: first substituting $p_2=s'+q_1+q_3$ in the total Wigner function, and then tracing out all variables except $s'$.  In this case, we get
\begin{equation}
\mathcal{P}_{A|D; \{1,2,3\}}(s,s') = \frac{1}{ \sqrt{\pi}\sigma} \exp\Big(-\frac{(s-s')^2}{\sigma^2} \Big)~,
\end{equation}
and thus
\begin{equation}
\mathcal{P}_{A;\{1,2,3\}}(s') = \frac{1}{\sqrt{\pi}\sqrt{\sigma^2 +\Sigma^2}} e^{-s'^2/(\sigma^2+\Sigma^2)}~.
\end{equation}

For the adversary structure parties 4 and 5, the reduced Wigner function of $W_{E|D;\{4,5\}}$ and $W_{E;\{4,5\}}$ can be obtained by tracing out the contribution of parties 1,2,3.  The covariance matrices of these states are
\begin{eqnarray}
\bm{V}_{E|D;\{4,5\}} = \left(\begin{array}{cccc}\frac{1}{2\sigma^2} & 0 & 0 & \frac{1}{2\sigma^2} \\0 & \frac{1}{\sigma^2}+\frac{\sigma^2}{2} & \frac{1}{2\sigma^2} & 0 \\0 & \frac{1}{2\sigma^2} & \frac{1}{2\sigma^2} & 0 \\\frac{1}{2\sigma^2} & 0 & 0 & \frac{1}{\sigma^2}+\frac{\sigma^2}{2}\end{array}\right), \\
\bm{V}_{E;\{4,5\}} = \left(\begin{array}{cccc}\frac{1}{2\sigma^2} & 0 & 0 & \frac{1}{2\sigma^2} \\0 & \frac{1}{\sigma^2}+\frac{\sigma^2+\Sigma^2}{2} & \frac{1}{2\sigma^2} & 0 \\0 & \frac{1}{2\sigma^2} & \frac{1}{2\sigma^2} & 0 \\\frac{1}{2\sigma^2} & 0 & 0 & \frac{1}{\sigma^2}+\frac{\sigma^2+\Sigma^2}{2}\end{array}\right),
\end{eqnarray}
where the symplectic spectrum are $\nu_{E|D;\{4,5\}}=\{\sqrt{1+\sigma^4}/2\sigma^2,\sqrt{1+\sigma^4}/2\sigma^2\}$ and $\nu_{E;\{4,5\}}=\{\sqrt{1+\sigma^4}/2\sigma^2,\sqrt{1+\sigma^4+2\sigma^2\Sigma^2}/2\sigma^2\}$.  The von Neumann entropy can then be calculated using Eq.~(\ref{eq:Neumann}), and the secret sharing rate is calculated from Eq.~(\ref{eq:Kcc}).

\subsubsection{Parties \{1,3,4\} collaborating}

Consider now that parties 1, 3, and 4 are the access structure.  In the infinitely squeezing case, the secret $s$ can be obtained when party 1 measures $\hat{p}$, party 3 and 4 measures $\hat{p}'=(\hat{p}-\hat{q})/\sqrt{2}$.  Because $\hat{N}_1-\hat{N}_3-\hat{N}_4$ is a nullifier, their measurement results are correlated as $-p_1+\sqrt{2}p'_3+\sqrt{2}p'_4=s$.
The reduced Wigner function of parties \{2,5\} collaboration is a constant function, so the secret sharing is secure.

In the finitely squeezing case, the probability distribution of the quantity $s'=-p_1+p_3-q_3+p_4-q_4$ can be obtained by first substituting the set of variables $(q_1,p_1,q_3,p_3,q_4,p_4)$ with the new set of variables $(q_1,s',q_3,p_3,q_4,p_4)$.  The determinant of the Jacobian matrix of this transformation is $1$.  All variables except $s'$ are traced out from the reduced Wigner function, then we get
\begin{equation}
\mathcal{P}_{A|D;\{1,3,4\}}(s,s')=\frac{1}{\sqrt{3\pi}\sigma} \exp\Big(-\frac{(s-s')^2}{3 \sigma^2}\Big)~,
\end{equation}
and thus
\begin{equation}
\mathcal{P}_{A;\{1,3,4\}}(s')= \frac{1}{\sqrt{\pi}\sqrt{3\sigma^2 +\Sigma^2}} e^{-s'^2/(3\sigma^2+\Sigma^2)}~.
\end{equation}

For the adversary structure parties 2 and 5, the covariance matrices of the states $\hat{\rho}_{E|D;\{2,5\}}$ and $\hat{\rho}_{E;\{2,5\}}$ are
\begin{eqnarray}
\bm{V}_{E|D;\{2,5\}} = \left(\begin{array}{cccc}\frac{1}{2\sigma^2} & 0 & 0 & 0 \\0 & \frac{1}{\sigma^2}+\frac{\sigma^2}{2} & 0 & \frac{1}{2\sigma^2} \\0 & 0 & \frac{1}{2\sigma^2} & 0 \\0 & \frac{1}{2\sigma^2} & 0 & \frac{1}{\sigma^2}+\frac{\sigma^2}{2} \end{array}\right),\\
\bm{V}_{E;\{2,5\}} = \left(\begin{array}{cccc}\frac{1}{2\sigma^2} & 0 & 0 & 0 \\0 & \frac{1}{\sigma^2}+\frac{\sigma^2}{2}+\frac{\Sigma^2}{2} & 0 & \frac{1}{2\sigma^2}+\frac{\Sigma^2}{2} \\0 & 0 & \frac{1}{2\sigma^2} & 0 \\0 & \frac{1}{2\sigma^2}+\frac{\Sigma^2}{2} & 0 & \frac{1}{\sigma^2}+\frac{\sigma^2}{2}+\frac{\Sigma^2}{2} \end{array}\right).
\end{eqnarray}
The symplectic spectrum are $\nu_{E|D;\{2,5\}}=\{\sqrt{1+\sigma^4}/2\sigma^2,\sqrt{3+\sigma^4}/2\sigma^2\}$ and $\nu_{E;\{2,5\}}=\{\sqrt{1+\sigma^4}/2\sigma^2,\sqrt{3+\sigma^4+2\sigma^2\Sigma^2}/2\sigma^2\}$, respectively.  The von Neumann entropy can then be calculated by Eq.~(\ref{eq:Neumann}), and hence the secret sharing rate by Eq.~(\ref{eq:Kcc}).


\subsubsection{Results}

\begin{figure}
\centering
\includegraphics{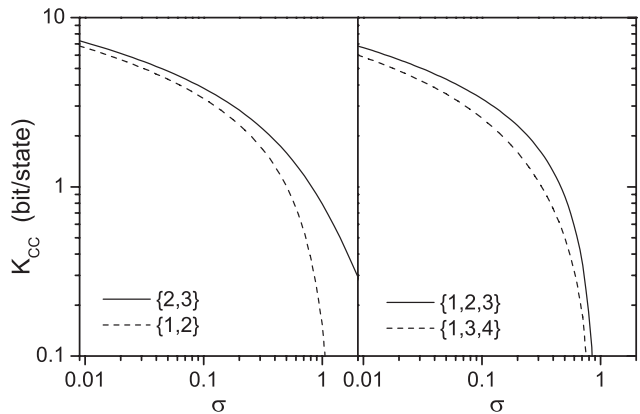}\caption{\label{fig:CC} Secret sharing rate of CC QSS protocols using CV cluster states with different squeezing parameters $\sigma$.  The variance of the classical secret probability is chosen as $\Sigma=1$.  Left panel: (2,3) protocol for \{2,3\} collaboration (solid line) and \{1,3\} collaboration (dashed line).  Left panel: (3,5) protocol for \{1,2,3\} collaboration (solid line) and \{1,3,4\} collaboration (dashed line). }
\end{figure}

The secret sharing rate for the (2,3)- and (3,5)-protocols is plotted against $\sigma$ in Fig~\ref{fig:CC}.  Apart from the $\{2,3\}$ in (2,3)-protocol that can completely remove the entanglement from the adversary structure, CC QSS is secure unless the squeezing parameter is larger than some threshold limit.  The threshold for both protocols we considered are around $\sigma\approx 1$.  This suggests a CC QSS can be implemented with modest technological requirement.

On the contrary to common beliefs that a CV state with $\sigma=1$ cannot transmit secure information, we note that the CC secret sharing rate is non-zero in some scenarios even when $\sigma\geq1$.  The result is not surprising in cluster state QSS, because implementing a \small CPHASE \normalsize requires the initial modes to be squeezed.  In fact, a two-mode cluster state can be easily showed to be local unitarily equivalent to a finitely-squeezed EPR state unless $\sigma \rightarrow \infty$.

Surprisingly, although any access structure collaboration can obtain $s$ in infinitely squeezing case, different collaborations obtain different security rate in the finitely squeezing case.  The unequal secret sharing rate is related to the entanglement structure of the cluster state.  In practice, the dealer has to consider the disadvantage of certain collaborations when applying CC QSS.

\section{CQ Quantum secret sharing\label{sec:CQ QSS}}

In the CQ setting of QSS, the dealer is connected to the parties through insecure quantum channels, so the unauthorised parities can manipulate all the modes sent from the dealer.  The CC protocol mentioned in Sec. \ref{sec:CC QSS} is insecure in this setting, because the adversary structure can capture and measure the modes to obtain $s$.  This eavesdropping can be intractable if the adversary structure resends to the access structure an infinitely squeezed state with the same $s$ encoded, so the access structure with have the same measurement result as the adversary structure.

Here we modify the CC protocol for the CQ setting.  We first present an entanglement-based protocol, and discuss how it can be reduced to a mixed-state protocol that reduces the resource requirement.  Instead of constructing an $n$-mode cluster state and encoding a classical secret $s$ into the state, the dealer prepares an $(n+1)$-mode standard cluster state, where $n$ of the modes are delivered to the parties while the dealer keeps the remaining one, denoted as mode $D$.  A good CQ protocol should produce quantum correlation between the dealer and the access structure much stronger than that between the dealer and the unauthorised parties.

Here we make two assumptions to simplify the demonstration of the security, but the assumptions will be relaxed at the end of this section without compromising the secret sharing rate.
We assume the access structure parties are connected by secure and ideal quantum channels, so the modes can be sent to one party, say party $h$, with perfect fidelity.
We also assume both the dealer and the access structure have quantum memories, so the cluster states in each round of CQ are stored with perfect fidelity for subsequent quantum operation and measurement.  

The procedure of CQ QSS is outlined as follows.  In each round of CQ QSS, an entangled state is shared between the dealer and the access structure.  Consider that the strong correlation is represented by the two nullifiers, $\hat{p}_D-\hat{Q}_A$ and $\hat{q}_D-\hat{P}_A$, which are linear combinations of the standard nullifiers.  $\hat{Q}_A$ and $\hat{P}_A$ are linear combinations of only the access structure parties' local $\hat{q}$ and $\hat{p}$ operators.  By applying a global operation, $\hat{U}_A$, on all the modes at party $h$, $\hat{Q}_A$ and $\hat{P}_A$  are transformed to $\hat{q}_h$ and $\hat{p}_h$.  As a result, the strong correlations with mode $D$ are transferred to mode $h$.

After all the rounds of cluster state distribution, the stage of parameter-estimation ensues.  The dealer or party $h$ randomly selects half of the shared modes for measurement, and the selection is announced.  Both the dealer and party $h$ measures the selected modes in either the $\hat{x}$ and $\hat{p}$ basis.  The measurement outcomes are announced for characterising the unmeasured states.


In the infinitely squeezing case, the estimated parameters should indicate that the state between mode $D$ and mode $h$ is maximally entangled. \footnote{In this paper, we refer to `maximally entangled' as an entangled state originating from two infinitely squeezed modes.}
The dealer and party $h$ measure each residual modes randomly in either the $\hat{q}$ or $\hat{p}$ basis, the basis is then announced.  Each measurement outcome is a random number on the real axis, and the outcomes are the same if the measurement bases are matching, i.e., one party measures in $\hat{q}$ while the other measures in $\hat{p}$.  The common random numbers can be used as secure keys to encode the secrets.
Because the state was maximally entangled, no information is leaked to unauthorised parties.

In the finitely squeezing case, although mode $D$ and mode $h$ are strongly correlated, the state of adversary structure is still weakly correlated with mode $D$.  As to be discussed, local quantum operation is applied on each residual mode to rectify the covariance of the state according to the estimated parameters.  The dealer and party $h$ then measure each mode in either the $\hat{x}$ or $\hat{p}$ basis, and announce the basis.
Unlike the infinitely squeezed case, post-processing is required to distill secure keys from the correlated measurement outcomes due to two reasons.  First, even if the measurement basis is matching, the measurement outcomes of the dealer and the access structure are merely strongly correlated but not exactly equal.  Besides, partial information about the outcomes is leaked to unauthorised parties due to the non-uniform distribution of the measurement outcomes and the weak entanglement between the unauthorised parties' and the dealer's modes.  

In the following, we employ the security analysis techniques from CV-QKD \cite{RaulThesis} to estimate the minimal rate of secure key distilled from each cluster state, and hence the secret sharing rate in each round of CQ QSS.

\subsection{Equivalence of CQ Quantum secret sharing and QKD}

We now show why CQ QSS and CV-QKD can be analysed using the same techniques.  Consider before the \small CPHASE \normalsize operation, mode $D$ is squeezed with $\sigma_D$ while all other modes are squeezed with $\sigma$.  Assume the mode $D$ is connected to $N$ neighbours after the cluster state formation, the reduced Wigner function of mode $D$ is
\begin{equation}\label{eq:WignerD}
W_D (q_D, p_D) = \frac{\sigma \sigma_D e^{-\sigma_D^2 q_D^2}}{\pi\sqrt{N+\sigma^2 \sigma_D^2}}\exp \Big(-\frac{p_D^2}{\sigma_D^2+N/\sigma^2} \Big) ~.
\end{equation}
$W_D$ is the same as the reduced Wigner function of a two-mode cluster state $|\mathcal{C}_N\rangle$, where mode $D$ is connected to a mode $u$ that is squeezed with $\sigma/\sqrt{N}$.  Because both the CQ cluster state and $|\mathcal{C}_N\rangle$ are pure, the amount of entanglement between the mode $D$ and the delivered modes is the same as the entanglement between the modes in $|\mathcal{C}_N\rangle$.  

As in the common security analysis of QKD, we grant the unauthorised parties the full power to manipulate the modes sent from the dealer.  Then there will be no difference for the dealer to prepare the CQ cluster state or $|\mathcal{C}_N\rangle$, because the unauthorised party can transform the delivered cluster state modes to mode $u$ or vice versa.  Then our CQ state delivery is equivalent to the following scenario: The dealer first prepares $|\mathcal{C}_N\rangle$ and delivers mode $u$ through an insecure quantum channel.  The unauthorised parties capture mode $u$, entangle it with ancillae, and forward some modes to the access structure.

\begin{figure}
\centering
\includegraphics{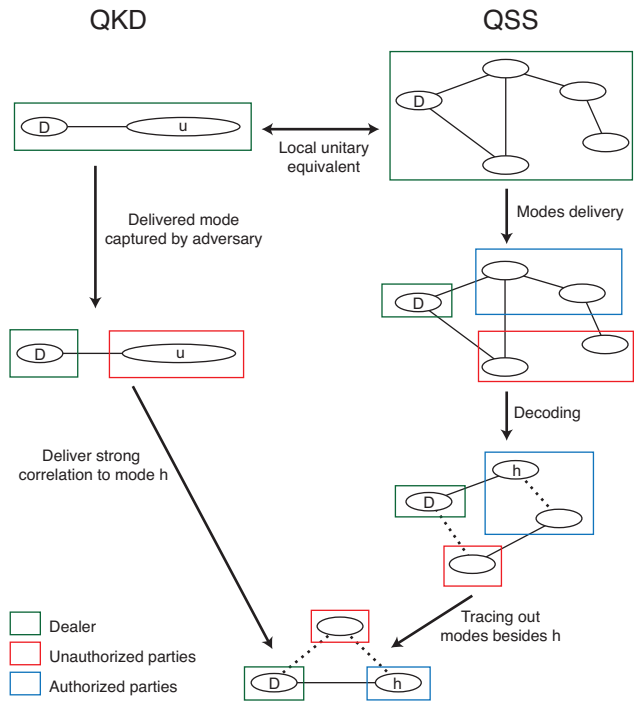}\caption{\label{fig:CQdeduce} Strategy for computing the secret sharing rate using CV-QKD techniques.  Strongly (weakly) correlated modes are linked by solid (dotted) lines.  The procedure of CQ QSS is shown on the right while that of QKD is shown on the left.  The key idea is that both QKD and QSS have the same initial (pure entangled state with parts delivered) and final resources (strongly correlated state between the dealer and the authorised parties.)}
\end{figure}

The access structure's modes are then gathered at party $h$.  After the operation $\hat{U}_A$, modes other than mode $h$ are still weakly correlated with mode $D$.  For simplicity, these weak correlations are neglected in our analysis, i.e., all modes except $h$ are traced out.  This action only reduces the quantum correlations between the dealer and the access structure, thus the security is not unphysically improved.  Now the CQ protocol is effectively reduced to a CV-QKD protocol: The dealer first prepares a two-mode Gaussian state, $|\mathcal{C}_N\rangle$, and delivers one mode.  The access structure finally gets a mode $h$ that remains strongly correlated with mode $D$, but the quantum correlation is reduced due to the entanglement with the environment controlled by the unauthorised parties.  The degradation of quantum correlations in the encoding and decoding processes in CQ QSS can be analogous to the loss and noise when transmitting an EPR state through an imperfect channel in QKD.  The whole idea is summarised schematically in Fig. \ref{fig:CQdeduce}.

\subsection{Secret sharing rate \label{subsec:ssr}}

In the unified picture of CV-QKD, a finitely squeezed two-mode squeezed state is prepared by the dealer and delivered to the authorised party through an imperfect channel \cite{GarciaPatron:2006p14390, Weedbrook:2012p13102}.  Both parties measure some of the delivered states to estimate the covariance matrix, $\bm{V}$, of the unmeasured states.  
Using the fact that Gaussian states minimise the distilled secure key rate for every state sharing the same $\bm{V}$ \cite{Wolf:2006p14389,GarciaPatron:2006p14390}, assuming the unmeasured states being Gaussian upper-bounds the information leakage to unauthorised parties.  
Because a Gaussian state is completely characterised by its covariance matrix, the secure key rate can be deduced from only $\bm{V}$.  
For realistic channels that are usually symmetrical for quadratures $\hat{x}$ and $\hat{p}$, $\bm{V}$ can be expressed in a standardised form as
\begin{equation}\label{eq:Vstandard}
\bm{V}=\left(\begin{array}{cc}V \bm{I} & c\bm{Z} \\c \bm{Z} & V'\bm{I} \end{array}\right)~,
\end{equation}
where $\bm{I}$ and $\bm{Z}$ are the $2\times 2$ identity and Pauli $Z$ matrices respectively; $V$ is the variance of the undelivered mode of the dealer; $V'$ is the variance of the mode received by the authorised party; $c$ is the correlation between the two modes.  $\bm{V}$ can be characterised by only $V$ and two channel parameters, the transmittance, $\tau$, and the noise, $\chi$, which are defined by the relations
\begin{equation}
c=\sqrt{\tau(V^2-\frac{1}{4})}~;~V'=\tau(V+\chi)~.
\end{equation}

To estimate the minimal secret sharing rate of the CQ QSS, in the parameter-estimation stage the dealer and the access structure construct the covariance matrix by measuring some of their modes.  A
quantum operation is then applied on the unmeasured modes to transform their covariance matrix to the standard form, from where
the analogous $\tau$ and $\chi$ can be extracted according to Eq.~(\ref{eq:Vstandard}).  For the pedagogic purpose, we demonstrate  in Sec. \ref{subsec:23CQ} and \ref{subsec:35CQ} the procedure of getting the standardised covariance matrix for different collaborations in the (2,3)- and (3,5)-CQ protocols respectively.  Readers who are mainly interested in the general formalism can skip the examples.
We assume our protocol is direct reconciliation, i.e., the measurement result of the dealer is the secret value that has to be estimated by the access structure, but the secret sharing rate of a reverse reconciliation protocol can be easily calculated by similar procedure \cite{RaulThesis}.  The secret sharing rate, $K_{CQ}$, is given by \cite{RaulThesis}
\begin{equation}\label{eq:CQrate}
K_{CQ}=I(D:A)-I(D:E)~,
\end{equation}
where $I(D:A)$ is the mutual information between the dealer and access structure; the information obtained by unauthorised parties 
is given by $I(D:E)$, which is capped by the Holevo bound.

The mutual information $I(D:A)$ can be calculated by comparing the variance of mode $h$ with and without knowing the measurement results of mode $D$.  In terms of the analogous channel parameters, the mutual information is given by \cite{RaulThesis}
\begin{equation}\label{eq:HDA}
I(D:A) = \frac{1}{2}\log \Big(\frac{V+\chi}{\chi +\frac{1}{4V}} \Big)~.
\end{equation}
In direct reconciliation protocols, the Holevo bound of the unauthorised parities' information is defined as
\begin{equation}\label{eq:HDE}
I(D:E)=S(E)-S(E|D)~,
\end{equation}
where $S(E)$ is the von Neumann entropy of unauthorised parties' state, $S(E|D)$ is the conditional von Neumann entropy if the measurement result of the dealer is known.  As the unauthorised parties can control the environment that purifies the whole system, the entropy of the unauthorised parties is the same as that of the system $DA$, i.e., $S(E)=S(DA)$.  The entropy can be calculated using Eq.~(\ref{eq:Neumann}) as $S(DA)=g(\nu_+)+g(\nu_-)$ \cite{Weedbrook:2012p13102}, where the symplectic spectrum of $\bm{V}$, $\{\nu_+,\nu_- \}$, is given by 
\begin{equation}\label{eq:nupm}
\nu_\pm=\frac{1}{2}\left( \sqrt{(V+V')^2-4 c^2} \pm (V-V') \right)~.
\end{equation}

Similarly, because the state of system $AE$ is pure after system $D$ is measured, the conditional entropy $S(E|D)$ is the same as $S(A|D)$.  The covariance matrix of system $A$ after the measurement of the dealer is given by \cite{Eisert:2002p13472,Fiurasek:2002p14342}
\begin{equation}
\bm{V}_{A|D} = \left(\begin{array}{cc}V-c^2/V & 0 \\0 & V'\end{array}\right)~,
\end{equation}
where the symplectic eigenvalue is 
\begin{equation}\label{eq:nuc}
\nu_c = \sqrt{V'\left(V'-\frac{c^2}{V}\right)} = \tau \sqrt{(V+\chi)\left(\frac{1}{V}+\chi\right) }~.
\end{equation}
Hence we get $S(E|D)=g(\nu_c)$.

With the strongly correlated measurement outcomes and the expected amount of secret information, $K_{CQ}$, the dealer and the access structure can collaborate to distill secure key to encode the classical secret \cite{VanAssche:2004p13958}.

\subsection{Example 1 of CQ Quantum secret sharing: (2,3)-protocol \label{subsec:23CQ}}

\begin{figure}
\centering
\includegraphics{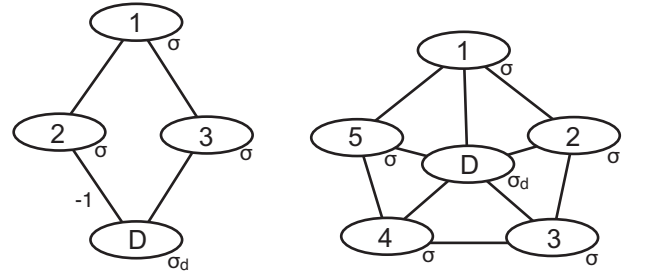}\caption{\label{fig:CQlayout} Schematic representation of the cluster state for the (2,3) CQ protocol (left), and the (3,5) CQ protocol (right).  All of the modes are not displaced before the \small CPHASE \normalsize operation.  The strength of unlabelled edges is $A=1$.}
\end{figure}

In a (2,3)-CQ protocol, any two of the three parties can form a strong correlation with the dealer, while any one party is only weakly correlated with the dealer.  The protocol can be implemented by a diamond-shaped CV cluster state with $A_{D3}=A_{13}=A_{12}=1$ and $A_{D2}=-1$, as shown in Fig. \ref{fig:CQlayout}.  We note that the diamond-shaped CV cluster state is also a form of error correction of a CV cluster state \cite{VanLoock:2007p13575}.  In the infinitely squeezing case, the nullifiers are
\begin{eqnarray}\label{eq:23CQ}
\hat{N}_D=\hat{p}_D+\hat{q}_2-\hat{q_3}&;&\hat{N}_1=\hat{p}_1-\hat{q}_2-\hat{q_3} \nonumber \\
\hat{N}_2=\hat{p}_2+\hat{q}_D-\hat{q_1}&;&\hat{N}_3=\hat{p}_3-\hat{q}_D-\hat{q_1}~.
\end{eqnarray}
The finitely squeezed state is described by Eq.~(\ref{eq:Wigner}) with the above nullifiers.

The access structure can be composed by parties $\{1,2\}$, $\{1,3\}$, and $\{2,3\}$.  The entanglement structure possessed by collaborations $\{1,2\}$ and $\{1,3\}$ are equivalent, because the nullifiers of $\{1,2\}$ will be the same as that of $\{1,3\}$ if the dealer applies a $\pi$-phase operation, $\hat{F}(\pi)$, to his mode.  On the other hand, the collaboration $\{2,3\}$ possesses a different entanglement structure.

\subsubsection{Parties \{1,2\} collaboration \label{subsubsec:12CQ}}

If parties $\{1,2\}$ are the access structure, the strong correlations are specified by the nullifiers 
\begin{eqnarray} \label{eq:nullifier12}
\hat{N}_D-\hat{N}_1=\hat{p}_D-\hat{p}_1+2\hat{q}_2\textrm{ and }
\hat{N}_2=\hat{q}_D-\hat{q}_1+\hat{p}_2~.
\end{eqnarray}
A global operation is applied on the access structure's mode to transfer the strong correlation to mode $2$, i.e., mode $2$ is treated as mode $h$.  The transformation can be implemented by various sequence of operations, but the final measurement results and the covariance matrix are not affected.  One possible choice is the \textit{\{1,2\} Decoding Sequence}: (i) apply $\exp(-i\hat{q}_1\hat{q}_2)$; (ii) then $\exp(i\hat{p}_1\hat{p}_2)$; (iii) finally $\hat{F}_2(\pi)$.  

After tracing out the modes other than mode $D$ and mode $2$, the covariance matrix of the resultant state $\hat{\rho}_{DA}$ is given by
\begin{equation}\label{eq:V12}
\bm{V}_{DA;\{1,2\}} = \left(\begin{array}{cccc}\frac{1}{2\sigma_D^2} & 0 & 0 & \frac{1}{2\sigma_D^2} \\0 & \frac{1}{\sigma^2}+\frac{\sigma_D^2}{2} & \frac{1}{\sigma^2} & 0 \\ 0 & \frac{1}{\sigma^2} &  \frac{1}{\sigma^2}+\frac{\sigma^2}{2}& 0 \\ \frac{1}{2\sigma_D^2} & 0 & 0 & \frac{\sigma^2}{2}+\frac{1}{2\sigma_D^2}  \end{array}\right)~.
\end{equation}
The covariance matrix will be revealed in the parameter-estimation stage when half of the states are measured.
As $\bm{V}_{DA;\{1,2\}}$ is not in the standard form, i.e., Eq.~(\ref{eq:Vstandard}), rectifying quantum operations are applied onto the residual states.  First of all, the variance of $\hat{q}_D$ and $\hat{p}_D$ are balanced by squeezing mode $D$ with the squeezing parameter
\begin{equation}\label{eq:gammad23}
\gamma_D=\sqrt{\frac{\sigma_D}{\sigma} \sqrt{2+\sigma^2\sigma_D^2}}~.
\end{equation}
Next, mode $2$ is squeezed to balance the coherent terms, i.e., $\langle \Delta\hat{q}_D\Delta\hat{p}_2 \rangle$ and $\langle \Delta\hat{p}_D\Delta\hat{q}_2 \rangle$.  
The squeezing parameter is given by
\begin{equation}
\gamma_2 = \sqrt{\frac{2 \sigma_D}{\sigma \sqrt{2+\sigma^2 \sigma_D^2}}}~.
\end{equation}
In practice, both $\gamma_D$ and $\gamma_2$ can be obtained from the results in parameter-estimation stage, i.e., without knowing the squeezing parameter of the initial cluster state.  This squeezing stage will transform the state $\hat{\rho}_{DA}$ as
\begin{equation}\label{eq:squeezingstage}
\hat{\rho}_{DA}\rightarrow \hat{\rho}'_{DA}= \hat{S}_D(\gamma_D)\hat{S}_2(\gamma_2) \hat{\rho}_{DA} \hat{S}_2^\dag(\gamma_2)\hat{S}_D^\dag(\gamma_D)~,
\end{equation}
where the covariance matrix becomes
\begin{equation}
\bm{V}'_{DA;\{1,2\}} = \left(\begin{array}{cccc}V_{(2,3)} & 0 & 0 & c \\0 & V_{(2,3)} & c & 0 \\0 & c & V'_q & 0 \\c & 0 & 0 & V'_p\end{array}\right)~,
\end{equation}
for $V_{(2,3)}= \sqrt{2+\sigma^2\sigma_D^2}/2\sigma\sigma_D$ ; $c=1/\sqrt{2}\sigma\sigma_D$; the variance of mode $2$ is
\begin{equation}\label{eq:Vqp12}
V'_q = \frac{1+\sigma^2\sigma_D^2}{\sigma\sigma_D \sqrt{2+\sigma^2\sigma_D^2}} ~;~
V'_p = \frac{(2+\sigma^4)\sqrt{2+\sigma^2\sigma_D^2}}{4 \sigma \sigma_D}~.
\end{equation}
We note that $\gamma_D$ and $V_{(2,3)}$ are the same for any collaboration in the (2,3)-protocol, because mode $D$ is kept with the dealer that is not affected by operations on delivered modes.
The disparity between $V'_q$ and $V'_p$ implies imbalanced noise for the quadratures $\hat{q}_2$ and $\hat{p}_2$.
As the aim of this paper is to demonstrate the \textit{possibility} of performing CQ QSS using CV cluster states, we balance the variances of $\hat{q}_2$ and $\hat{p}_2$ by a `state-averaging' process, which would nonetheless sacrifice some quantum correlations.  
Consider the dealer randomly divides the unmeasured states into two sets, and the choice of division is announced.  In one set, the dealer applies a Fourier operator, $\hat{F}_D(-\pi/2)$, on each mode that transforms the quadrature operators as $\hat{q}_D\rightarrow \hat{p}_D$ and $\hat{p}_D\rightarrow -\hat{q}_D$.  In the other set, party $2$ applies $\hat{F}_D(-\pi/2)$ on each mode that causes the transformation $\hat{q}_2\rightarrow \hat{p}_2$ and $\hat{p}_2\rightarrow -\hat{q}_2$.  After that, the choice of division is discarded.  The state will be transformed as
\begin{eqnarray}\label{eq:stateaverage}
\hat{\rho}'_{DA} \rightarrow \hat{\rho}''_{DA} &=& \frac{1}{2}\hat{F}_D(-\pi/2)\hat{\rho}'_{DA}\hat{F}^\dag_D(-\pi/2) \nonumber \\
&&+ \frac{1}{2} \hat{F}_2(-\pi/2)\hat{\rho}'_{DA}\hat{F}^\dag_2(-\pi/2)~.
\end{eqnarray}
The covariance matrix of $\hat{\rho}''_{DA}$ is given by
\begin{equation}\label{eq:Vt12}
\bm{V}''_{DA;\{1,2\}}= \left(\begin{array}{cccc}V_{(2,3)} & 0 & c & 0 \\0 & V_{(2,3)} & 0 & -c \\ c & 0 & V_{A;\{1,2\}} & 0 \\0 & -c & 0 & V_{A;\{1,2\}}\end{array}\right)~,
\end{equation}
where $V_{A;\{1,2\}}=(V'_q+V'_p)/2$ with the definition in Eq.~(\ref{eq:Vqp12}).  Finally, the modes are measured by the dealer and party $2$ in either the $\hat{q}$ or $\hat{p}$ basis.  The variances of the measurement results will be given by Eq.~(\ref{eq:Vt12}).

We note that $\hat{\rho}''_{DA}$ is not a Gaussian state because the stage-averaging process in Eq.~(\ref{eq:stateaverage}) is not a Gaussian operation.  
To calculate the secret sharing rate using the techniques in Sec. \ref{subsec:ssr}, we assume the measurement results originates from a Gaussian state $\hat{\rho}_G$ where its covariance matrix is $\bm{V}$.  According to Ref.s \cite{Wolf:2006p14389,GarciaPatron:2006p14390}, Gaussian states minimise the secure key rate for all states with the same covariance matrix.  Therefore our action maximises the power of the unauthorised parties and lower-bounds the secret sharing rate.

By comparing Eq.~(\ref{eq:Vt12}) with Eq.~(\ref{eq:Vstandard}), the variance of the dealer's mode is $V=V_{(2,3)}$, and the analogous channel parameters can be deduced as $\tau=c^2/(V^2_{(2,3)}-1/4)=1$ and $\chi=V_{A;\{1,2\}}-V_{(2,3)}$.  The minimal secret sharing rate can then be calculated by Eqs.~(\ref{eq:CQrate}-\ref{eq:nupm}) and (\ref{eq:nuc}).

\subsubsection{Parties \{2,3\} collaboration \label{subsubsec:23CQ}}

If parties $\{2,3\}$ are the access structure, the strong correlations are specified by the nullifiers 
\begin{equation}\label{eq:nullifier23}
\hat{N}_D=\hat{p}_D+\hat{q}_2-\hat{q}_3\textrm{ and }\frac{\hat{N}_2 - \hat{N}_3}{2}=\hat{q}_D+\frac{\hat{p}_2}{2}-\frac{\hat{p}_3}{2}~.
\end{equation}
The quantum correlation can be transferred to mode $2$ by the \textit{\{2,3\} Decoding Sequence}, which is simply  a 50:50 beam splitter that transforms $\hat{a}_2\rightarrow -(\hat{a}_2+\hat{a}_3)/\sqrt{2}$ and $\hat{a}_3\rightarrow (\hat{a}_2-\hat{a}_3)/\sqrt{2}$.  The resultant covariance matrix between mode $D$ and mode $2$ becomes
\begin{equation}\label{eq:V23}
\bm{V}_{DA;\{2,3\}} = \left(\begin{array}{cccc}\frac{1}{2 \sigma_D^2} & 0 & 0 & \frac{1}{\sqrt{2}\sigma_D^2} \\0 & \frac{1}{\sigma^2}+\frac{\sigma_D^2}{2} & \frac{1}{\sqrt{2}\sigma^2} & 0 \\0 & \frac{1}{\sqrt{2}\sigma^2} & \frac{1}{2\sigma^2} & 0 \\\frac{1}{\sqrt{2}\sigma_D^2} & 0 & 0 & \frac{\sigma^2}{2}+\frac{1}{\sigma_D^2}\end{array}\right)~.
\end{equation}


The secure sharing rate can be deduced by similar process as in the $\{1,2\}$ collaboration: squeezing and transforming local modes to construct a state with standardised covariance matrix, and then measuring the states to obtain the analogous channel parameters for computing the information of different parties.  However, the $\{2,3 \}$ collaboration is special as the dealer and party $2$ are actually holding a pure state, i.e., the beam splitter has removed all entanglement from the unauthorised parties.  This can be seen from the symplectic spectrum of Eq.~(\ref{eq:V23}), $\nu_{\{2,3\}} = \{1/2,1/2 \}$, so the entropy of the system $DA$ vanishes, i.e., $S(DA)=0$.  Therefore the unauthorised parties cannot obtain any information about the secret by entangling their modes to the dealer's mode.  The secret sharing rate is hence the same as the mutual information between the dealer and the access structure, which is given by
\begin{equation}
I(D:A)=\log(2 V_{(2,3)})~.
\end{equation}

\subsection{Example 2 of CQ Quantum secret sharing: (3,5)-protocol \label{subsec:35CQ}}

In a (3,5)-CQ protocol, any three of the five parties can form a strong correlation with the dealer, while any collaboration with less than two parties is only weakly correlated with the dealer.  The protocol can be implemented by a pentagonal CV cluster state, as shown in Fig. \ref{fig:CQlayout}, where each connected vertex is entangled by a \small CPHASE \normalsize operation with $\mathcal{A}_{ij}=1$.  In the infinitely squeezing case, the nullifiers are
\begin{eqnarray}\label{eq:35CQ}
\hat{N}_D=\hat{p}_D-\sum_{i=1}^5 \hat{q}_i~;~\hat{N}_i = \hat{p}_i-\hat{q}_{i+1}-\hat{q}_{i-1}-\hat{q}_D~,
\end{eqnarray}
where $i+1=1$ when $i=5$; $i-1=5$ when $i=1$.  The finitely squeezed state is described by the Wigner function in Eq.~(\ref{eq:Wigner}) with the above nullifiers.

The access structure can be composed by two categories of collaboration: three neighbouring parties, e.g. parties $\{1,2,3 \}$, and two neighbours with one disjoint party, e.g. $\{1,3,4\}$.  The collaborations in each category hold nullifiers with the same form, so the decoding sequence will be the same.  If the squeezing parameter is identical for all five modes, the secret sharing rate of the collaborations in each category will also be the same.


\subsubsection{Parties \{1,2,3\} collaboration \label{subsubsec:123CQ}}

If parties $\{1,2,3\}$ are the access structure, the strong correlations are specified by the nullifiers 
\begin{eqnarray}
\hat{N}_D-\hat{N}_1+2\hat{N}_2-\hat{N}_3&=&\hat{p}_D-(\hat{p}_1+3\hat{q}_1)+(2\hat{p}_2+\hat{q}_2)  \nonumber \\
&&-(\hat{p}_3+3\hat{q}_3) \nonumber \\
\textrm{and }-\hat{N}_2&=&\hat{q}_D - \hat{p}_2 + \hat{q}_1 + \hat{q}_3~.\label{eq:nullifier123}
\end{eqnarray}
The quantum correlations can be transferred to party $2$ by the \textit{\{1,2,3\} Decoding Sequence}: (i) applying $\exp(-i\hat{q}_1\hat{q}_2)$ and $\exp(-i\hat{q}_2\hat{q}_3)$; (ii) then $\exp(i\hat{p}_1\hat{p}_2)$ and $\exp(i\hat{p}_2\hat{p}_3)$; (iii) finally $\exp(i\hat{p}_2(2\hat{p}_1+\hat{q}_1))$ and $\exp(i\hat{p}_2(2\hat{p}_3+\hat{q}_3))$.  

The covariance matrix of the state $\hat{\rho}_{DA}$ between mode $D$ and mode $2$ is given by
\begin{equation}\label{eq:V123}
\bm{V}_{DA;\{1,2,3\}} = 
\left(\begin{array}{cccc}\frac{1}{2\sigma_D^2} & 0 & 0 & \frac{1}{2\sigma_D^2} \\0 & \frac{5+\sigma^2\sigma_D^2}{2\sigma^2} & \frac{5}{2\sigma^2} & 0 \\0 & \frac{5}{2\sigma^2} & \frac{5+6\sigma^4}{2\sigma^2} & -\sigma^2 \\\frac{1}{2\sigma_D^2} & 0 & -\sigma^2 & \frac{1+\sigma^2\sigma_D^2}{2\sigma_D^2}\end{array}\right)~.
\end{equation}
All terms in $\bm{V}_{DA;\{1,2,3\}}$ can be revealed by $\hat{x}$ and $\hat{p}$ measurements in the parameter-estimation stage except for the local coherent terms $\langle(\Delta \hat{q}_2 \Delta\hat{p}_2+\Delta \hat{p}_2 \Delta\hat{q}_2)/2 \rangle$.  However these terms do not affect the parameters in the squeezing stage, and will be eventually cancelled during state-averaging.


The unmeasured states are squeezed locally to balance the variance of $\hat{q}_D$ and $\hat{p}_D$, as well as the coherent terms.  The state is transformed as in Eq.~(\ref{eq:squeezingstage}), where the parameters for the $\{1,2,3\}$ collaborations are
\begin{equation}
\gamma_D = \sqrt{ \frac{\sigma_D}{\sigma} \sqrt{5+\sigma^2\sigma_D^2}}~;~
\gamma_2 = \sqrt{\frac{\sigma}{5\sigma_D}\sqrt{5+\sigma^2\sigma_D^2}}~.
\end{equation}
The covariance matrix of the transformed state is given by
\begin{equation}
\bm{V}'_{DA;\{1,2,3\}} = \left(\begin{array}{cccc}V_{(3,5)} & 0 & 0 & c \\0 & V_{(3,5)} & c & 0 \\0 & c & V'_q & 0 \\c & 0 & 0 & V'_p\end{array}\right)
\end{equation}
where $V_{(3,5)}= \sqrt{5+\sigma^2\sigma_D^2}/2\sigma\sigma_D$ ; $c=\sqrt{5}/2\sigma\sigma_D$, and the variance of mode $2$ is given by
\begin{equation}\label{eq:Vqp123}
V'_q = \frac{(5+6\sigma^4)\sqrt{5+\sigma^2\sigma_D^2}}{10\sigma\sigma_D}~;~V'_q = \frac{5(1+\sigma^2\sigma_D^2)}{2\sigma\sigma_D \sqrt{5+\sigma^2\sigma_D^2}}~.
\end{equation}
The value of $\gamma_D$ and $V_{(3,5)}$ are the same for any collaboration in the (3,5)-protocol.

State-averaging ensues to balance the correlations of $p_D-q_2$ and $q_D-p_2$.  Half of the unmeasured states are transformed by $\hat{F}_D(-\pi/2)$, while the other half are transformed by $\hat{F}_2(-\pi/2)$.  After discarding the choice of division, the state transforms as Eq.~(\ref{eq:stateaverage}), where
the covariance matrix becomes
\begin{equation}\label{eq:Vt123}
\bm{V}''_{DA;\{1,2,3\}}= \left(\begin{array}{cccc}V_{(3,5)} & 0 & c & 0 \\0 & V_{(3,5)} & 0 & -c \\ c & 0 & V_{A;\{1,2,3\}} & 0 \\0 & -c & 0 & V_{A;\{1,2,3\}}\end{array}\right)~,
\end{equation}
for $V_{A;\{1,2,3\}}=(V'_q+V'_p)/2$ with the definition in Eq.~(\ref{eq:Vqp123}).  We note that the local coherent terms vanish after state-averaging because their sign in $\hat{F}_D(-\pi/2)\hat{\rho}'_{DA}\hat{F}^\dag_D(-\pi/2)$ and $\hat{F}_2(-\pi/2)\hat{\rho}'_{DA}\hat{F}^\dag_2(-\pi/2)$ are opposite.

We again assume the measurement results come from a Gaussian state with the same covariance matrix $\bm{V}''_{DA;\{1,2,3\}}$.  The variance of the dealer's mode is recognised as $V=V_{(3,5)}$, and the analogous channel parameters can be deduced as $\tau=\bar{c}^2/(V^2_{(3,5)}-1/4)=1$ and $\chi=V_{A;\{1,2,3\}}-V_{(3,5)}$.  The minimal secret sharing rate can then be calculated by Eqs.~(\ref{eq:CQrate}-\ref{eq:nupm}) and (\ref{eq:nuc}).

\subsubsection{Parties \{1,3,4\} collaboration \label{subsubsec:134CQ}}

If parties $\{1,3,4\}$ are the access structure, the strong correlations are specified by the nullifiers 
\begin{eqnarray}
\hat{N}_D-2\hat{N}_1+\hat{N}_3+\hat{N}_4&=&\hat{p}_D-(2\hat{p}_1+\hat{q}_1)+(\hat{p}_3-2\hat{q}_3) \nonumber \\
&&+(\hat{p}_4-2\hat{q}_4) \nonumber \\
\textrm{and }\hat{N}_1-\hat{N}_3-\hat{N}_4&=&\hat{q}_D+\hat{p}_1-(\hat{p}_3-\hat{q}_3) \nonumber \\
&&-(\hat{p}_4-\hat{q}_4)~.\label{eq:nullifier134}
\end{eqnarray}
The quantum correlations can be transferred to mode $1$ by the \textit{\{1,3,4\} Decoding Sequence}: (i) apply $\exp(-i(\hat{q}_1+\hat{p}_3-\hat{q}_3))$ and $\exp(-i(\hat{q}_1+\hat{p}_4-\hat{q}_4))$; (ii)  followed by $\exp(i\hat{p}_1\hat{p}_3)$ and $\exp(i\hat{p}_1\hat{p}_4)$; (iii) then $\exp(i2\hat{p}_1^2)$; (iv) finally $\hat{F}(\pi)$.  

The covariance matrix of the state $\hat{\rho}_{DA}$ between mode $D$ and mode $1$ is given by
\begin{equation}\label{eq:V134}
\bm{V}_{DA;\{1,3,4\}} = 
\left(\begin{array}{cccc}\frac{1}{2\sigma_D^2} & 0 & 0 & \frac{1}{2\sigma_D^2} \\0 & \frac{5+\sigma^2\sigma_D^2}{2\sigma^2} & \frac{5}{2\sigma^2} & 0 \\0 & \frac{5}{2\sigma^2} & \frac{5+6\sigma^4}{2\sigma^2} & -2\sigma^2 \\ \frac{1}{2\sigma_D^2} & 0 & -2\sigma^2 & \frac{1+3\sigma^2\sigma_D^2}{2\sigma_D^2}\end{array}\right)~.
\end{equation}

After the parameter-estimation stage, local-squeezing is applied as in Eq.~(\ref{eq:squeezingstage}) except mode $1$ is now mode $h$.
The squeezing parameter in the current collaboration is
\begin{equation}
\gamma_D = \sqrt{\frac{\sigma_D}{\sigma} \sqrt{5+\sigma^2\sigma_D^2}}~;~
\gamma_1 = \sqrt{\frac{\sigma}{5\sigma_D} \sqrt{5+\sigma^2 \sigma_D^2} }~.
\end{equation}
The covariance matrix of the unmeasured states then becomes
\begin{equation}
\bm{V}'_{DA;\{1,3,4\}} = \left(\begin{array}{cccc}V_{(3,5)} & 0 & 0 & c \\ 0 & V_{(3,5)} & c & 0 \\0 & c & V'_q & 0 \\ c & 0 & 0 & V'_p\end{array}\right)
\end{equation}
where $c=\sqrt{5}/2\sigma\sigma_D$, and the variance of mode $1$ is given by
\begin{equation}\label{eq:Vqp134}
V'_q = \frac{(5+6\sigma^4)\sqrt{5+\sigma^2\sigma_D^2}}{10 \sigma \sigma_D}~;~
V'_p = \frac{5(1+3\sigma^2\sigma_D^2)}{2\sigma\sigma_D \sqrt{5+\sigma^2\sigma_D^2}}~.
\end{equation}

State-averaging ensues to balance the correlations of $p_D-q_1$ and $q_D-p_1$.  
The covariance matrix becomes
\begin{equation}\label{eq:Vt134}
\bm{V}''_{DA;\{1,3,4\}}= \left(\begin{array}{cccc}V_{(3,5)} & 0 & c & 0 \\0 & V_{(3,5)} & 0 & -c \\ c & 0 & V_{A;\{1,3,4\}} & 0 \\0 & -c & 0 & V_{A;\{1,3,4\}}\end{array}\right)~,
\end{equation}
where $V_{A;\{1,3,4\}}=(V'_q+V'_p)/2$ with the definition in Eq.~(\ref{eq:Vqp134}).  Similar to the $\{1,2,3\}$ collaboration, the local coherent terms are eliminated by state-averaging.

After local $\hat{q}$ and $\hat{p}$ measurements, we again assume the results come from a Gaussian state.  The variance of the dealer's mode is recognized as $V=V_{(3,5)}$, and the analogous channel parameters can be deduced as $\tau=\bar{c}^2/(V^2_{(3,5)}-1/4)=1$ and $\chi=V_{A;\{1,3,4\}}-V_{(3,5)}$. 

\subsection{Results}

\begin{figure}
\centering
\includegraphics{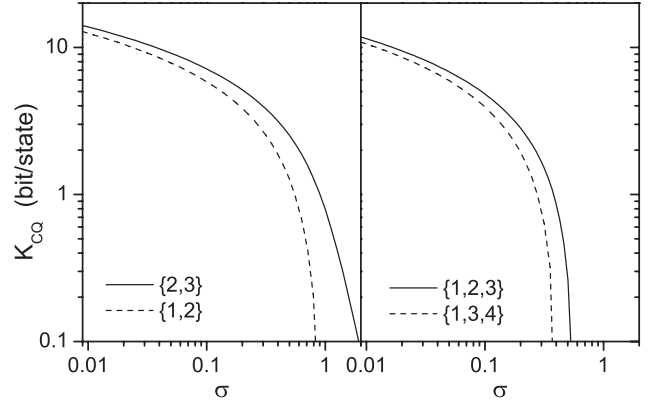}\caption{\label{fig:CQ} Secret sharing rate of CQ QSS protocols using CV cluster states with different squeezing parameters $\sigma$.  The squeezing parameter of dealer's mode is set as $\sigma_D=\sigma$.  Left panel: (2,3) protocol for \{2,3\} collaboration (solid line) and \{1,3\} collaboration (dashed line).  Right panel: (3,5) protocol for \{1,2,3\} collaboration (solid line) and \{1,3,4\} collaboration (dashed line). }
\end{figure}

The secret sharing rate of the (2,3)- and (3,5)-CQ protocol is plotted in Fig. \ref{fig:CQ} for different $\sigma$.  As in the CC case, secure key can be distilled if the squeezing parameter is smaller than a threshold limit, except in the $\{2,3\}$ collaboration in (2,3)-protocol that the entanglement of the adversary structure is completely removed.  The threshold values are about $\sigma\approx 1$ in the examples we consider.  Disparity of secure key rate in different collaborations is also observed in CQ protocols.  The secret sharing rate is non-zero in some cases even when $\sigma \geq 1$.  As we have discussed in the CC case, a two-mode cluster state is still entangled even if the initial state is not squeezed.  The entanglement between the dealer and party $h$ imposes strong quantum correlations that can distill secure keys.

\subsection{Simplification}

In the above analysis of CQ protocols, we assume an $(n+1)$-mode cluster state has to be created and quantum memory is available to store the delivered modes until all of the states are received as well as an ideal quantum channel being available between the access structure parties.  Here we show that these requirements can be relaxed without compromising the security.

\subsubsection{Mixed state approach}

Because the distributed modes are unaffected by any local operation of the dealer, the state obtained by the parties is the same regardless of whether mode $D$ has been measured.  Therefore, instead of preparing an $(n+1)$-mode cluster state $|\Psi\rangle$ and measuring mode $D$ afterwards, the dealer can simulate the consequence of the measurement by distributing an $n$-mode state that is the same as the measured $|\Psi\rangle$.  Consider if the dealer intends to measure in the $\hat{q}$ basis, then the dealer can instead prepare the pure state $(\langle s |_{q_D})|\Psi\rangle$, where $|s\rangle_{q_D}$ is a $\hat{q}_D$ eigenstate with the eigenvalue $s$.  Similarly in the $\hat{p}$ measurement rounds, the dealer can prepare $(\langle s |_{p_D})|\Psi\rangle$, where $|s\rangle_{p_D}$ is the $\hat{p}_D$ eigenstate with the eigenvalue $s$.  Other parties cannot distinguish the mixed state from $|\Psi\rangle$ if $s$ is picked according to the probability distributions of dealer's measurement results on $|\Psi\rangle$.  The probability distributions are imposed by the Wigner function in Eq.~(\ref{eq:WignerD}) as
\begin{eqnarray}
\mathcal{P}_{q_D}(s) = \int W_D(s,p_D) dp_D = \frac{\sigma_D}{\sqrt{\pi}} e^{-\sigma_D^2 s^2}~; \\
\mathcal{P}_{p_D}(s) = \int W_D(q_D,s) dq_D = \frac{e^{-\frac{s^2}{\sigma_D^2+N/\sigma^2}}}{\sqrt{\pi}\sqrt{\sigma_D^2+N/\sigma^2}} ~.
\end{eqnarray}

In the simulated $\hat{q}$ measurement rounds, the infinitely squeezed $(\langle s |_{q_D})|\Psi\rangle$ is characterised by the nullifiers
\begin{equation}
\hat{N}_i^q = \hat{p}_i - \sum_{j \in \mathcal{N}} \mathcal{A}_{ij} \hat{q}_j - A_{i D}s~,
\end{equation}
where $i=1,\ldots,n$; $A_{iD}=0$ if the mode $i$ is not a neighbour of mode $D$.  The nullifiers $\hat{N}_i^q$ is the same as $\hat{N}_i$ except the operator $\hat{q}_D$ is replaced by the simulated measurement outcome $s$.  
In the finitely squeezed case, the state can be characterised by the Wigner function $W_{q_D}(q_1,p_1,\ldots,q_n,p_n)$, which is obtained by tracing out $p_D$ and replacing all $q_D$ by $s$ in the Wigner function of $|\Psi\rangle$.  Because $W_{q_D}$ is the same as $W_c$ in Eq.~(\ref{eq:Wigner}) with the nullifiers $\hat{N}_i^q$, $(\langle s |_{q_D})|\Psi\rangle$ is a cluster-class state that can be formed by displacing a finitely squeezed cluster state.

In the simulated $\hat{p}$ measurement rounds, the infinitely squeezed $(\langle s |_{p_D})|\Psi\rangle$ is characterised by the nullifiers
\begin{equation}
\hat{N}_1^p =  \sum_{j \in \mathcal{N}} A_{jD} \hat{q}_j - s~;~\hat{N}_i^p=\hat{N}_i-\hat{N}_{i-1}~,
\end{equation}
where $j=2,\ldots,n$.  In the finitely squeezed case, the state can be characterised by the Wigner function $W_{p_D}(q_1,p_1,\ldots,q_n,p_n)$, which is obtained by tracing out $q_D$ and replacing all $p_D$ by $s$  in the Wigner function of $|\Psi\rangle$.  However unlike $W_{q_D}$, $W_{p_D}$ is not necessarily representable by the form of $W_c$ with the nullifiers $\hat{N}_1^p$, therefore $(\langle s |_{p_D})|\Psi\rangle$ is generally not a cluster-class state.  Nevertheless, $(\langle s |_{p_D})|\Psi\rangle$ is a Gaussian state that can be efficiently prepared by offline squeezed states and linear optical elements \cite{VanLoock:2007p13575}.

\subsubsection{Classical memory}

When estimating the secret sharing rate, we need the delivered modes to be rectified so the covariance matrix is in the standard form.  The modes are stored in quantum memories until the covariance matrix is constructed from parameter-estimation.  
Here we show that the measurement probability distribution of the transformed state $\hat{\rho}''$ can be obtained by: first measuring the original state $\hat{\rho}$, and then subjecting the measurement results to classical manipulations.  Therefore the delivered modes can be measured before the parameter-estimation stage, quantum memory is thus not necessary.

The rectifying process involves two stages: local-squeezing and state-averaging.  After state-averaging, $\hat{\rho}''$ becomes a mixture of $\hat{F}_D(-\pi/2)\hat{\rho}'\hat{F}^\dag_D(-\pi/2)$ and $\hat{F}_h(-\pi/2)\hat{\rho}'\hat{F}^\dag_h(-\pi/2)$.  By definition, the Wigner function of $\hat{\rho}''$, $W''$, can be written as the sum of the Wigner function of $\hat{\rho}'$, $W'$, as
\begin{eqnarray}
W''(q_D,p_D,q_h,p_h) &=& \frac{1}{2} W'(p_D,-q_D,q_h,p_h) \nonumber \\
&&+ \frac{1}{2} W'(q_D,p_D,p_h,-q_h)~.
\end{eqnarray}

Consider the dealer measures $\hat{\rho}''$ in $\hat{q}_D$ and party $h$ measures in $\hat{q}_h$, the probability of obtaining measurement outcomes $y_1$ and $y_2$, $\mathcal{P}''_{q_D,q_h}(y_1, y_2)$, is given by
\begin{eqnarray}\label{eq:mix}
\mathcal{P}''_{q_D,q_h}(y_1, y_2) &=& \int W''(y_1,p_D,y_2,p_h) dp_D dp_h \nonumber \\
&=& \frac{1}{2} \int W'(p_D,-y_1,y_2,p_h) dp_D dp_h \nonumber \\
&&+ \frac{1}{2} \int W'(y_1,p_D,p_h,-y_2) dp_D dp_h \nonumber \\
&=&\frac{1}{2} \mathcal{P}'_{p_D, q_h} (-y_1,y_2) + \frac{1}{2} \mathcal{P}'_{q_D,p_h}(y_1,-y_2) ~,
\nonumber \\
\end{eqnarray}
where the last equality involves renaming of variables; $\mathcal{P}'_{x_D,x_h}$ is the joint $\{\hat{x}_D, \hat{x}_h \}$ measurement probability of $\hat{\rho}'$.  Similarly, the probability of another strongly correlated measurement, $\{\hat{p}_D, \hat{p}_h\}$, can be expressed as
\begin{equation}
\mathcal{P}''_{p_D,p_h}(y_1, y_2)=\frac{1}{2} \mathcal{P}'_{q_D, p_h} (y_1,y_2) + \frac{1}{2} \mathcal{P}'_{p_D,q_h}(y_1,y_2) ~.
\end{equation}

These two relations indicate that the measurement probability distributions \textit{after} state-averaging are not different from mixing some measurement probability distributions \textit{before} state-averaging.  Consider the dealer and party $h$ randomly measure $\hat{\rho}'$ in $\hat{x}$ and $\hat{p}$ basis.  Half of the $\{\hat{q}_D, \hat{p}_h\}$ outcomes and half of the $\{\hat{p}_D, \hat{q}_h\}$ outcomes are picked to mimic the $\{\hat{q}_D, \hat{q}_h\}$ measurement of $\hat{\rho}''$.  For the $\{\hat{q}_D, \hat{p}_h\}$ half, all the $\hat{p}_h$ outcomes are multiplied by $-1$ and then regarded as $\hat{q}_h$ outcomes; for the $\{\hat{p}_D, \hat{q}_h\}$ half, all the $\hat{p}_D$ outcomes are multiplied by $-1$ and then regarded as $\hat{q}_D$ outcomes.  After mixing the two sets of data, the probability $\mathcal{P}^m$ of getting $y_1, y_2$ is given by
\begin{equation}
\mathcal{P}^m(y_1,y_2)=\frac{1}{2} \mathcal{P}'_{p_D, q_h} (-y_1,y_2) + \frac{1}{2} \mathcal{P}'_{q_D,p_h}(y_1,-y_2) ~,
\end{equation}
which is the same as Eq.~(\ref{eq:mix}).  The $\{\hat{p}_D, \hat{p}_h\}$ measurement probability of $\hat{\rho}''$ can be mimicked by similar procedures.  

In local-squeezing stage, the dealer and party $h$ apply local squeezing operations, $\hat{S}_D(\gamma_D)$ and $\hat{S}_h(\gamma_h)$ , to balance the variance of mode $D$ and the coherent terms.  The Wigner function of the state $\hat{\rho}$ transforms as
\begin{eqnarray}
W(q_D,p_D,q_h,p_h)&\rightarrow& W'(q_D,p_D,q_h,p_h) \nonumber \\
&=& W\left(\frac{q_D}{\gamma_D},\gamma_D p_D,\frac{q_h}{\gamma_h},\gamma_h p_h\right)~.
\end{eqnarray}

Consider $\hat{\rho}'$ is measured in $\hat{q}_D$ and $\hat{p}_h$ basis, the probability of obtaining the outcomes $y_1, y_2$ is given by
\begin{eqnarray}\label{eq:saqp}
\mathcal{P}_{q_D,p_h} (y_1,y_2) &\rightarrow& \mathcal{P}'_{q_D,p_h} (y_1,y_2) \nonumber \\
&=& \int W\left(\frac{y_1}{\gamma_D},\gamma_D p_D,\frac{q_h}{\gamma_h},\gamma_h y_2\right) dp_D dq_h \nonumber \\
&=& \frac{\gamma_h}{\gamma_D} \mathcal{P}_{q_D,p_h} \left(\frac{y_1}{\gamma_D},\gamma_h y_2\right)~,
\end{eqnarray}
where $\mathcal{P}$ is the probability distribution when measuring $\hat{\rho}$.  Similarly, the probability distribution of $\hat{p}_D$ and $\hat{q}_h$ measurement transforms as
\begin{equation}\label{eq:sapq}
\mathcal{P}_{p_D,q_h} (y_1,y_2) \rightarrow \frac{\gamma_D}{\gamma_h} \mathcal{P}_{p_D, q_h} (\gamma_D y_1,\frac{y_2}{\gamma_h})~.
\end{equation}
Measurement results of $\{\hat{q}_D,\hat{q}_h\}$ and $\{\hat{p}_D,\hat{p}_h\}$ are sifted as they are merely weakly correlated.

In fact, physically squeezing the state is not necessary because the transformations in Eqs.~(\ref{eq:saqp}) and (\ref{eq:sapq}) can be conducted by classically scaling the measurement results.
Consider every $\hat{q}_i$ measurement results are scaled by $1/\gamma_i$, and every $\hat{p}_i$ measurement results are scaled by $\gamma_i$.  The old probability, $\mathcal{P}$, of a $\hat{q}$ measurement result lying in the range $[y,y+d y]$, is equal to the new probability, $\mathcal{P}^s$, of a scaled result in the range of $[\gamma y,\gamma y+\gamma dy]$.  Thus we have the relation $\mathcal{P}(y)dy=\mathcal{P}'(\gamma y) \gamma dy$ for $\hat{q}$ measurement, and similarly $\mathcal{P}(y)dy=\mathcal{P}'(y/\gamma) dy/\gamma$ for $\hat{p}$ measurement.
By eliminating common factors and redefining variables, we get
\begin{eqnarray}
\mathcal{P}^s_{q_D,p_h} (y_1,y_2)=\frac{\gamma_h}{\gamma_D} \mathcal{P}_{q_D,p_h} \left(\frac{y_1}{\gamma_D},\gamma_h y_2\right)~;\nonumber \\
\mathcal{P}^s_{p_D,q_h} (y_1,y_2)=\frac{\gamma_D}{\gamma_h} \mathcal{P}_{p_D, q_h} \left(\gamma_D y_1,\frac{y_2}{\gamma_h}\right)~.
\end{eqnarray}
The above resultant probability distributions are the same as Eq.~(\ref{eq:saqp}) and (\ref{eq:sapq}).

\subsubsection{Local measurement}

We have assumed the access structure parties have forwarded their modes to a single party for global operations.  Here we show that the measurement results of the dealer and access structure remains strongly correlated even if the access structure conducts local measurements only.

Recall the strong correlation is represented by the nullifiers $\hat{p}_D-\hat{Q}_A$ and $\hat{q}_D-\hat{P}_A$.  Because they are linear combinations of standard nullifiers, both the operators $\hat{Q}_A$ and $\hat{P}_A$ are sums of local operators, i.e.,
\begin{equation}\label{eq:QPA}
\hat{Q}_A=\sum_j^n k_j^q \hat{M}_j^q~;~\hat{P}_A=\sum_j^n k_j^p \hat{M}_j^p~,
\end{equation}
where $k_j^q$ and $k_j^p$ are real coefficients; $\hat{M}_j^q$ and $\hat{M}_j^p$ are rotated quadrature operators of mode $j$.  After discussing through secure classical channels to decide the measurement basis to be $\hat{Q}_A$ or $\hat{P}_A$, the dealer and authorised parties homodyne detect their modes according to the basis $\hat{M}_j^q$ or $\hat{M}_j^p$.  The measurement results are then shared among access structure through secure classical channels.

Without loss of generality, we consider that the access structure has chosen to measure $\hat{Q}_A$.  The measurement result $Q_A$ is a linear combination of local measurement results $M_j^q$, i.e., $Q_A = \sum_j^n k_j^q M_j^q$. 
The strong correlation is observed from the joint probability distribution of $p_D$ and $Q_A$, which we will show is the same as the joint probability distribution of $p_D$ and $q_h$.
Consider the Wigner function of the state of the dealer and the access structure, $W_{DA}(q_D,p_D,\bm{q}_A,\bm{p}_A)$ where $\bm{q}_A$ and $\bm{p}_A$ are the quadrature variables of the access structure, is obtained by tracing out the unauthorised parties' contributions in $W_c$ in Eq.~(\ref{eq:Wigner}).
When rewritten in terms of the new variables $\bm{M}^q=\{M_j^q\}$ and $\bm{*M}^q=\{*M_j^q\}$, the Wigner function becomes
\begin{equation}\label{eq:Wignert1}
W_{DA}(q_D,p_D,\bm{q}_A,\bm{p}_A) \equiv W'_{DA}(q_D,p_D,\bm{M}^q,\bm{*M}^q)~,
\end{equation}
where $*M_j^q$ is the complementary variable of $M_j^q$, i.e., the corresponding operators satisfy $[\hat{M}_j^q, *\hat{M}_j^q]=i$.  The choice of $\{*\hat{M}_j^q\}$ is not unique, but we can pick the set that $\hat{P}_A$ is a linear combination of.  

We construct another set of variables $\bm{Q}=\{Q_A,Q_2,\ldots,Q_m\}$ and $\bm{P}=\{P_A,P_2,\ldots,P_m\}$, where $\bm{Q}$ ($\bm{P}$) involves linear combinations of $\{M_j^q\}$ ($\{*M_j^q\}$) only; and the corresponding operators obey the commutation relations: $[\hat{Q}_j, \hat{P}_l ]=i \delta_{jl}$, $[\hat{Q}_j, \hat{Q}_l ]=0$, and $[\hat{P}_j, \hat{P}_l ]=0$.  Such a construction of variables is possible as there exists unitary operators that transform $\hat{\bm{M}}^q$ to $\hat{\bm{Q}}$  and $\hat{\bm{*M}}^q$ to $\hat{\bm{P}}$ while preserving the commutation relations.  In terms of $\bm{Q}$ and $\bm{P}$, the Wigner function can be further rewritten as
\begin{equation}\label{eq:Wignert2}
W'_{DA}(q_D,p_D,\bm{M}^q,\bm{*M}^q) \equiv W''_{DA}(q_D,p_D,\bm{Q},\bm{P})~.
\end{equation}

The local measurement results follow a classical probability distribution $\mathcal{P}''_{DA}$, which is obtained by tracing out the complementary components, i.e.,
\begin{eqnarray}
\mathcal{P}''_{DA}(p_D, \bm{M}^q) &=& \int W'_{DA} dq_Dd^m (\bm{*M}^q_1) \nonumber \\
&=& \int W''_{DA} dq_Dd^m\bm{P} ~,
\end{eqnarray}
where the last equality is imposed because $\bm{P}$ is a linear combination of $\bm{*M}^q_1$ only.  The probability distribution of $Q_A$ is obtained by tracing out the other independent variables in $\mathcal{P}''_{DA}$, i.e.,
\begin{equation}\label{eq:Pclassical}
\mathcal{P}_{DA}(p_D, Q_A)=\int \mathcal{P}''_{DA} dQ_2 dQ_3 \dots~.
\end{equation}

On the other hand, consider the access structure parties' modes are transferred to party $h$.  The strong quantum correlation is transferred to mode $h$ by applying a global operation, $\hat{U}_A$, which transforms $\hat{Q}_A \rightarrow \hat{U}_A\hat{Q}_A\hat{U}^\dag_A = \hat{q}_h$ and $\hat{P}_A \rightarrow \hat{U}_A\hat{P}_A\hat{U}^\dag_A = \hat{p}_h$.  Other operators are transformed as $\hat{Q}_j \rightarrow\hat{y}_j$ and $\hat{P}_j\rightarrow\hat{z}_j$.  Using the definition in Eqs.~(\ref{eq:Wignert1}) and (\ref{eq:Wignert2}), the Wigner function becomes
\begin{equation}
W_{DA}(q_D,p_D,\bm{q},\bm{p})\rightarrow W''_{DA}(q_D,p_D,q_h,p_h,\bm{y},\bm{z})~,
\end{equation}
where $\bm{y}=(y_2,\ldots,y_m)^T$ and $\bm{z}=(z_2,\ldots,z_m)^T$.  Because $\hat{Q}_j$ and $\hat{P}_j$ commute with both $\hat{Q}_A$ and $\hat{P}_A$, the transformed operators $\hat{y}_j$ and $\hat{z}_j$ do not contain any attributes of mode $h$.  The joint probability distribution of $\hat{p}_D$ and $\hat{q}_h$ is obtained from the Wigner function after tracing out the modes other than mode $D$ and $h$, as well as the complementary variables $q_D$ and $p_h$, i.e.,
\begin{equation}\label{eq:Pquantum}
\mathcal{P}_{DA}(p_D,q_h)=\int W''_{DA} dq_Ddp_hd^{m-1}\bm{y}d^{m-1}\bm{z}~.
\end{equation}


The probability distribution in Eqs.~(\ref{eq:Pclassical}) and (\ref{eq:Pquantum}) are deduced by different procedures.  The former one is deduced by first obtaining the classical probability distribution of all local measurements, and then extracting the probability distribution of the classical variable $Q_A$; the later one is deduced by first achieving the Wigner function of the transformed quantum state, and then obtaining the measurement probability of the operator $\hat{q}_h$.  However, Eqs.~(\ref{eq:Pclassical}) and (\ref{eq:Pquantum}) are mathematically equivalent because their overall derivations are the same: tracing out all quadrature variables in the Wigner function except those specifying the strong correlation.  Similar analysis can be applied to the correlation between $q_D$ and $P_A$.  As a result, the access structure can obtain the same covariance matrix as we have discussed previously.  Hence the secret sharing rate remains unchanged even if the access structure's modes are measured locally.


\subsubsection{Simplified CQ protocol}

Incorporating the above ideas, the CQ protocol can be simplified as follow: an $(n+1)$ cluster state or an $n$ mode mixed state is prepared by the dealer and delivered.  Parties in the access structure have agreed on the measurement basis in each round, local measurements are conducted on each received modes.  The classical measurement results are shared among the access structure through secure classical channels.  Both the dealer and the access structure announce half of the results to estimate the covariance matrix; while the other half is scaled and mixed so the covariance matrix is in the standard form.  The variance and the analogous channel parameters are then recognised for calculating the secret sharing rate.  Finally a secure key is distilled from the strongly correlated measurement results, the key is then used for sharing classical secrets.

We end this section with two comments.  First, although the quantum channels for delivering cluster states are assumed to be ideal, we believe a modified version of our protocol would allow CQ QSS with realistic (lossy and noisy) channels.  The covariance matrix of the delivered modes can still be obtained by parameter-estimation, and subsequent classical manipulations can always turn the measurement results to obey the standard covariance matrix.  

Second, in all the examples we have investigated, the entanglement with the unauthorised parties' modes only add noise to the access structure parties' modes, but the analogous transmittance are retained as $1$.  This result is surprising in the scenario of CV-QKD, because imperfection is always simulated by adding noise into a beam splitter which reduces transmittance.  We believe this phenomenon originates from the distinctive entanglement structure in the resource states between CQ QSS (a cluster state) and CV-QKD (an EPR state).

\section{QQ Quantum secret sharing \label{sec:QQ QSS}}

In the QQ setting of QSS, the dealer shares a secret quantum state among parties by delivering a multipartite entangled state.  The channels connecting the dealer and the parties can be insecure, so the unauthorised parties can manipulate all the delivered states.  In an ideal QQ protocol, the access structure can recover the secret state with perfect fidelity, while the unauthorised parties cannot get any information about the state due to the quantum no-cloning theorem \cite{Scarani:2005p14399}.


Our QQ QSS scheme is a generalisation of the CQ protocol.  The dealer prepares an $(n+1)$ mode cluster state, of which $n$ modes are distributed to the parties while one is kept by the dealer.  After forming the collaborations, the access structure parties forward their modes to party $h$.   We assume the parties are connected by secure quantum channels, so the access structure can combine their modes without being eavesdropped.  A global operation is applied to extract a strongly entangled state between mode $D$ and a single mode $h$. 

For an infinitely squeezed QQ cluster state, the strong correlation between the dealer and the access structure is represented by the nullifiers $\hat{p}_D-\hat{Q}_A$ and $\hat{q}_D-\hat{P}_A$.  After all modes are gathered in party $h$, an operation $\hat{U}_A$ is applied to transfer the quantum correlation to mode $h$, i.e., $\hat{Q}_A\rightarrow\hat{U}_A \hat{Q}_A \hat{U}_A^\dag=\hat{q}_h$ and $\hat{P}_A\rightarrow\hat{U}_A \hat{P}_A \hat{U}_A^\dag=\hat{p}_h$.  We note that both the nullifiers, $\hat{Q}_A$ and $\hat{P}_A$, and the operation $\hat{U}_A$ are the same as that in the corresponding CQ protocol in Sec. \ref{sec:CQ QSS}.

The transformed nullifiers, $\hat{p}_D-\hat{q}_h$ and $\hat{q}_D-\hat{p}_h$, indicate that the dealer and party $h$ are sharing an infinitely squeezed two-mode cluster state, which is a CV maximally entangled state (for infinite energy).  By jointly measuring the secret state and the two-mode cluster, the dealer can teleport the secret state to mode $h$.  After appropriate error correction according to the dealer's measurement results, party $h$ can revert the secret state with perfect fidelity.

In the finitely squeezing case, party $h$ conducts the same $\hat{U}_A$ to transform the strong correlation to mode $h$.  However, mode $D$ and mode $h$ are not maximally entangled, because their state is finitely squeezed and is weakly entangled to other modes.  Conducting teleportation using the non-maximal entanglement will reduce the fidelity of the teleported state.  The inaccurately shared secret state may indicate a reduction of security of the QQ QSS, because some information about the secret state would be leaked through the measurement results announced by the dealer, and through the states held by the adversary structure that are weakly entangled with the teleported state.  

Instead of conducting teleportation after each round of QQ QSS, we consider the extracted state is stored in quantum memories.  After rounds of the QQ protocol, a more entangled state can be distilled from the stored extracted states through CV entanglement distillation \cite{Fiurasek:2003p14365,PhysRevA.67.062320,Eisert:2004p14366,Menzies:2007p14367,RalphLund:2009}.  Although distilling a maximally entangled CV state is impossible due to the infinite required energy, the enrichment of entanglement can enhance the fidelity of the teleportation.

The amount of entanglement of the distilled state is determined by that of each extracted state, as well as the number of extracted states accumulated in the quantum memory.  We quantify the amount of entanglement by the logarithmic negativity $\mathbb{E}$ \cite{Vidal:2002p13909}, which is the upper bound of the distillable entanglement.  The logarithmic negativity of a state $\hat{\rho}$ is defined as 
\begin{equation}
\mathbb{E}(\hat{\rho}) = \log|| \hat{\rho}^{T_A}||_1~,
\end{equation}
where the superscript $T_A$ denotes a partial transpose of the density matrix; $||\cdot||_1$ is the trace norm.  If $\hat{\rho}$ is a two-mode Gaussian state with a covariance matrix $\bm{V}$, the logarithmic negativity can be calculated as
\begin{equation}\label{eq:lognegativity}
\mathbb{E}(\hat{\rho}) = \sum_k F(\tilde{\nu}_k)~,
\end{equation}
where $F(x)=-\log(2x)$ if $x<1/2$, and $F(x)=0$ if $x\geq1/2$; $\{\tilde{\nu}_k\}$ is the symplectic spectrum of $\tilde{\bm{V}}$, which is defined as \cite{Braunstein:2005p9702}
\begin{equation}
\tilde{\bm{V}}=\left(\begin{array}{cc}\bm{I} & 0 \\0 & \bm{Z}\end{array}\right)\cdot\bm{V}\cdot\left(\begin{array}{cc}\bm{I} & 0 \\0 & \bm{Z}\end{array}\right)~.
\end{equation}
The covariance matrix $\bm{V}$ can be obtained by randomly measuring some of the stored states in either $\hat{q}$ or $\hat{p}$.

Because logarithmic negativity is additive \cite{Vidal:2002p13909}, at least $\mathbb{E}_0/\mathbb{E}$ extracted states with logarithmic negativity $\mathbb{E}$ is required to distill a two-mode squeezed vacuum state with logarithmic negativity $\mathbb{E}_0$.  In this section, we demonstrate the procedure of extraction, and calculate the logarithmic negativity of the extracted state in each round of the (2,3)- and the (3,5)-protocols.

We note that logarithmic negativity is additive but not strongly superadditive \cite{Wolf:2006p14389}, so the amount of entanglement may be overestimated if the access structure's modes in different rounds are entangled \cite{Plbnio:2007:IEM:2011706.2011707}, i.e., when the unauthorised parties conduct coherent attacks on the delivered modes.  In that case, the amount of entanglement should be characterised by other strongly superadditive entanglement measures, such as distillable entanglement and squashed entanglement \cite{Wolf:2006p14389}.  However, logarithmic negativity is applicable in the current case because of our assumption of ideal quantum channels, so the access structure is expected to get the same states as prepared by the dealer.  The adversary structure parties only get information about the shared secret through their modes obtained in each round, which is effectively a collective attack.

\subsection{Example 1 of QQ Quantum secret sharing: (2,3)-protocol}

In the (2,3) QQ protocol, any two out of the three participating parties can recover the shared secret state with high fidelity, while any one party achieves much less information about the secret.  This protocol can be implemented by the same diamond-shaped CV cluster state as that for the (2,3)-CQ protocol in Sec. \ref{subsec:23CQ}.  In the infinitely squeezing case, the nullifiers are given by Eq.~(\ref{eq:23CQ}), and these nullifiers with Eq.~(\ref{eq:Wigner}) characterise the finitely squeezed cluster state.

Three different collaborations can be formed: parties 1 and 2, parties 1 and 3, and parties 2 and 3.  The entanglement structure of the state of $\{1,2\}$ collaboration is the same as that of $\{1,3\}$ collaboration, as their states are equivalent up to a local unitary.  As a result, the entanglement extracted between the dealer and parties $\{1,2\}$ is the same as that between the dealer and parties $\{1,3\}$.


For the $\{1,2\}$ collaboration, the quantum correlation can be transferred to mode 2 by the \textit{\{1,2\} Decoding Sequence} in Sec \ref{subsubsec:12CQ}.  
In the infinitely squeezing case, the nullifiers in Eq.~(\ref{eq:nullifier12}), which specifies the strong correlation, is transformed to $\hat{p}_D-\hat{q}_2$ and $\hat{q}_D-\hat{p}_2$.  An infinitely squeezed two-mode cluster is hence extracted for teleportation.  In the finitely squeezing case, the \textit{\{1,2\} Decoding Sequence} also transfers the strong correlation to party 2.  The covariance matrix of the extracted state between mode $D$ and mode 2 is given by $\bm{V}_{DA;\{1,2\}}$ in Eq.~(\ref{eq:V12}).  


For the $\{2,3\}$ collaboration, the quantum correlation, which is specified by the nullifiers in Eq.~(\ref{eq:nullifier23}), can be transferred to party 2 by applying a 50:50 beam splitter between mode 2 and mode 3. 
An infinitely squeezed two-mode cluster state between mode $D$ and mode 2 is extracted in the infinitely squeezing case.  While in the finitely squeezing case, a strongly entangled state is extracted with the covariance matrix $\bm{V}_{DA;\{2,3\}}$ in Eq.~(\ref{eq:V12}).

\subsection{Example 2 of QQ Quantum secret sharing: (3,5)-protocol}

In the (3,5)-QQ protocol, any three out of the five participating parties can recover the shared secret state with high fidelity, while fewer than three parties achieve much less information about the secret.  This protocol can be implemented by the same pentagonal CV cluster state as used for the (3,5)-CQ protocol in Section \ref{subsec:35CQ}.  In the infinitely squeezing case, the nullifiers is given by Eq.~(\ref{eq:35CQ}), and these nullifiers with Eq.~(\ref{eq:Wigner}) characterise the finitely squeezed cluster state.  
Two categories of collaborations can be formed: three neighbouring parties, and two neighbours with one disjoint party.  Within each category, the procedure of decoding and the final entanglement extracted are the same for each collaboration.

Take the parties $\{1,2,3\}$ as an example of the three neighbouring parties collaboration.  The quantum correlation can be transferred to party 2 by the \textit{\{1,2,3\} Decoding Sequence} in Sec \ref{subsubsec:123CQ}. 
In the infinitely squeezing case, the nullifiers in Eq.~(\ref{eq:nullifier123}) that specify the strong correlation are transformed to $\hat{p}_D-\hat{q}_2$ and $\hat{q}_D-\hat{p}_2$.  This indicates an infinitely squeezed two-mode cluster is extracted in mode $D$ and mode 2.  While in the finitely squeezing case, the sequence of operations extracts a strongly entangled state with the covariance matrix $\bm{V}_{DA;\{1,2,3\}}$ in Eq.~(\ref{eq:V123}).  

On the other hand, parties $\{1,3,4\}$ is an example of the two neighbours with one disjoint party collaboration.  The quantum correlations can be transferred to party 1 by the \textit{\{1,3,4\} Decoding Sequence} in Sec \ref{subsubsec:134CQ}.  In the infinitely squeezing case, an infinitely squeezed two-mode cluster is extracted in mode $D$ and mode 1, because the nullifiers in Eq.~(\ref{eq:nullifier134}) that specify the strong correlation is transformed to $\hat{p}_D-\hat{q}_1$ and $\hat{q}_D-\hat{p}_1$.  While in the finitely squeezing case, a strongly entangled state is extracted, and its covariance matrix is given by $\bm{V}_{DA;\{1,3,4\}}$ in Eq.~(\ref{eq:V134}).  

\begin{figure}
\centering
\includegraphics{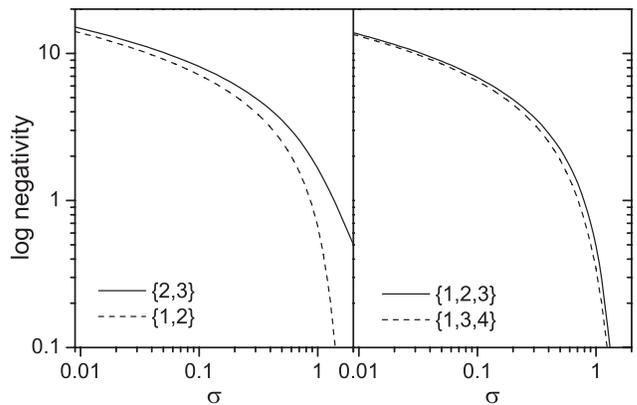}\caption{\label{fig:QQ} Logarithmic negativity of the state extracted from a CV cluster state in QQ QSS.  Left panel: (2,3)-protocol for \{2,3\} collaboration (solid line) and \{1,3\} collaboration (dashed line).  Right panel: (3,5)-protocol for \{1,2,3\} collaboration (solid line) and \{1,3,4\} collaboration (dashed line). }
\end{figure}

The logarithmic negativity of the extracted state for different collaborations in the (2,3)- and the (3,5)-protocols is calculated using Eq.~(\ref{eq:lognegativity}) with the corresponding covariance matrices.  The result is plotted in Fig. \ref{fig:QQ} against different squeezing parameters.

\section{conclusion \label{sec:conclusion}}

In this work, we extended the unified cluster state quantum secret sharing framework proposed in \cite{Markham:2008p13316,Keet:2010p13444} into the continuous-variable regime.  We proposed that all three tasks of quantum secret sharing can be implemented by CV cluster states.  Although a QQ protocol can be used to conduct CC and CQ, simplifications in the later two scenarios can reduce the requirement of resources.  For a CC protocol involving $n$ parties, only $n$-mode cluster states are needed, and the states can be measured once received.  A CQ protocol requires either a mixture of two $n$-mode Gaussian states or an $(n+1)$-mode cluster state.  The states can be locally measured once it is received.  A QQ protocol requires $(n+1)$-mode cluster states.  The states have to be transferred to one party and accumulated in quantum memories for entanglement distillation.

On the contrary to discrete-variable systems, where no known physical principle hinders the creation of a maximally entangled state, the creation of a maximally entangled CV state requires infinite energy and is thus not practical.  Finitely squeezed states are realistic substitutes for the maximally entangled resources, but the non-maximal entanglement would leak information about the shared secret to the unauthorised parties.
We proposed computable measures to account for the security of each of the three tasks of quantum secret sharing.  The secret sharing rate of a CC protocol is the difference between the mutual information between the dealer and the access structure, and the adversary structure's information that is capped by the Holevo bound.  The secret sharing rate of a CQ protocol can be computed by the secure key rate of the analogous QKD protocol.  The performance of a QQ protocol can be determined by the amount of extracted entanglement between the dealer and the access structure.

Although we have analysed only the (2,3)- and the (3,5)- protocols that are both threshold protocols \cite{Cleve:1999p5862}, our technique is applicable to non-threshold protocols because the security analysis involves only the variance of measurement results of the dealer and the access structure.  Security of general continuous-variable CQ and QQ protocols can also be analysed using our techniques, i.e., transferring the strong correlation to mode $h$ and then compute the covariance matrix between mode $D$ and mode $h$, even if the resource state is not a continuous-variable cluster state.


To the best of our knowledge, our work is the first one showing that quantum secret sharing is feasible with finitely squeezed CV resources.  Since a finitely squeezed cluster state can be deterministically constructed using only offline squeezers and linear optics, which are practically available resources in current technology, we believe CV quantum secret sharing can be implemented in the near future.  An important remaining question is to determine if the performance of quantum secret sharing is seriously worsened under the presence of environmental and apparatus noise.  As we borrow the security analysis techniques from QKD, which works well in noisy circumstances, it is likely that realistic performance of quantum secret sharing can be analysed using the formalisms presented in this work.

\begin{acknowledgements}
We thank Hoi-Kwong Lo and Barry Sanders for useful discussions.  This work is supported by NSERC and Canada Research Chairs program.
\end{acknowledgements}


\bibliographystyle{phaip}
\pagestyle{plain}
\bibliography{CVQSS_bib}

\end{document}